%% file: main.tex
\documentclass[sigconf]{acmart}

\AtBeginDocument{%
  \providecommand\BibTeX{{%
    \normalfont B\kern-0.5em{\scshape i\kern-0.25em b}\kern-0.8em\TeX}}}

\copyrightyear{2021}
\acmYear{2021}
\setcopyright{acmcopyright}\acmConference[KDD '21]{Proceedings of the 27th ACM SIGKDD Conference on Knowledge Discovery and Data Mining}{August 14--18, 2021}{Virtual Event, Singapore}
\acmBooktitle{Proceedings of the 27th ACM SIGKDD Conference on Knowledge Discovery and Data Mining (KDD '21), August 14--18, 2021, Virtual Event, Singapore}
\acmPrice{15.00}
\acmDOI{10.1145/3447548.3467186}
\acmISBN{978-1-4503-8332-5/21/08}
\usepackage{amsmath}

\usepackage[linesnumbered,ruled,vlined]{algorithm2e}
\usepackage{graphicx}
\usepackage{subcaption}
\usepackage{multirow}
\newtheorem{theorem}{Theorem}
\newtheorem{lemma}[theorem]{Lemma}
\usepackage{hhline}
\usepackage{hyperref}
\begin{document}
\fancyhead{}

%\title{MoCL: Contrastive Learning on Molecular Graphs with Multi-level Domain Knowledge}
\title{MoCL: Data-driven Molecular Fingerprint via Knowledge-aware Contrastive Learning from Molecular Graph}

\author{Mengying Sun}
\email{sunmeng2@msu.edu}
\affiliation{%
	\institution{Michigan State University}
	\city{East Lansing}
	\state{Michigan}
	\country{USA}
}
\author{Jing Xing}
\email{xingjin1@msu.edu}
\affiliation{%
	\institution{Michigan State University}
	\city{Grand Rapids}
	\state{Michigan}
	\country{USA}
}
\author{Huijun Wang}
\email{huijun.wang@agios.com}
\affiliation{%
	\institution{Agios Pharmaceuticals}
	\city{Cambridge}
	\state{Massachusetts}
	\country{USA}
}
\author{Bin Chen}
\email{chenbi12@msu.edu}
\affiliation{%
	\institution{Michigan State University}
	\city{Grand Rapids}
	\state{Michigan}
	\country{USA}
}
\author{Jiayu Zhou}
\email{jiayuz@msu.edu}
\affiliation{%
	\institution{Michigan State University}
	\city{East Lansing}
	\state{Michigan}
	\country{USA}
}

\renewcommand{\shortauthors}{Sun, et al.}

\begin{abstract}
    Recent years have seen a rapid growth of utilizing graph neural networks (GNNs) in the biomedical domain for tackling drug-related problems. However, like any other deep architectures, GNNs are data hungry. While requiring labels in real world is often expensive, pretraining GNNs in an unsupervised manner has been actively explored. Among them, graph contrastive learning, by maximizing the mutual information between paired graph augmentations, has been shown to be effective on various downstream tasks. However, the current graph contrastive learning framework has two limitations. First, the augmentations are designed for general graphs and thus may not be suitable or powerful enough for certain domains. Second, the contrastive scheme only learns representations that are invariant to local perturbations and thus does not consider the global structure of the dataset, which may also be useful for downstream tasks. In this paper, we study graph contrastive learning designed specifically for the biomedical domain, where molecular graphs are present. We propose a novel framework called MoCL, which utilizes domain knowledge at both local- and global-level to assist representation learning. The local-level domain knowledge guides the augmentation process such that variation is introduced without changing graph semantics. The global-level knowledge encodes the similarity information between graphs in the entire dataset and helps to learn representations with richer semantics. The entire model is learned through a double contrast objective. We evaluate MoCL on various molecular datasets under both linear and semi-supervised settings and results show that MoCL achieves state-of-the-art performance.
 
\end{abstract}



\maketitle

\section{Introduction}\label{sec:intro}
\input{intro}

\section{Related Work}\label{sec:related_work}
\input{related_work}

\section{Method}\label{sec:method}
\input{method}

\section{Experiment}\label{sec:exp}
\input{exp}

% emphasize esemble is on the data selection, not prediction, prediction only use the first networks for all methods.

\section{Conclusion}
In this work, we propose to utilize multi-level domain knowledge to assist the contrastive representation learning on molecular graphs. The local-level domain knowledge enables new augmentation scheme and global-level domain knowledge incorporates global structure of the data into the learning process. We demonstrate that both knowledge improve the quality of the learned representations.

\subsection*{Acknowledgment}
This research is funded in part by National Science Foundation under grant IIS-1749940 (JZ), Office of Naval Research under grant N00014-20-1-2382 (JZ), National Institute of Health under grants 1R01GM134307 (JZ, BC) and K01ES028047 (BC).

\bibliographystyle{ACM-Reference-Format}
\bibliography{reference}

\clearpage
\section*{Appendix}

\subsection*{Implementation Details}
Table \ref{detail} shows the detailed parameter settings for all datasets. Semi-ratio depends on the data size such that around 100 molecule labels are sampled from each dataset. The neighbor size also depends on the data size such that the number of clusters is between 5 and 30 for all datasets. The parameter $\lambda$ which controls the weight between local and global loss, and augmentation time for MoCL-DK are all set to the same set of values for all datasets.

\begin{table}[!h]
	\small \centering
	\setlength\tabcolsep{3.6pt} % default value: 6pt
	\begin{tabular}{cccccc}
		\hline
		Dataset & Size & Semi-ratio & Neigbor Size          & $\lambda$            & DK          \\ \hline
		bace    & 1513 & 0.05       & \{50, 100, 150, 300\} & \{0.5, 1, 5, 10\} & \{1,2,3,5\} \\
		bbbp    & 2050 & 0.05       & \{50, 100, 150, 300\} & \{0.5, 1, 5, 10\} & \{1,2,3,5\} \\
		clintox & 1483 & 0.05       & \{50, 100, 150, 300\} & \{0.5, 1, 5, 10\} & \{1,2,3,5\} \\
		mutag   & 188  & 0.5        & \{10, 20, 30, 40\}    & \{0.5, 1, 5, 10\} & \{1,2,3,5\} \\
		sider   & 1427 & 0.05       & \{50, 100, 150, 300\} & \{0.5, 1, 5, 10\} & \{1,2,3,5\} \\
		tox21   & 7831 & 0.01       & (600, 800, 1000\}     & \{0.5, 1, 5, 10\} & \{1,2,3,5\} \\
		toxcast & 8597 & 0.01       & \{600, 800, 1000\}    & \{0.5, 1, 5, 10\} & \{1,2,3,5\} \\ \hline
	\end{tabular}
	\caption{Detailed experimental settings for each dataset.}\label{detail}
\end{table}
\vspace{-2em}

\noindent Unlike prior work \cite{you2020graph} in which only node, node features and connectivity information are used as input, our GNN incorporates edge features, therefore, the implementation of general augmentation is slightly different from \cite{you2020graph}. We list the operations for both node (features) and edge (features) in Table \ref{app-aug}.   
\begin{table}[!h]
		\small \centering
	\setlength\tabcolsep{3.6pt} % default value: 6pt
	\begin{tabular}{ccccc}
		\hline
		Augmentation         & Node    & Node features & Edge     & Edge features \\ \hline
		Drop Node       & removed & removed       & removed  & removed       \\
		Perturb Edge    & -       & -             & permuted & permuted      \\
		Subgraph        & subsample    & subsample          & keep     & keep          \\
		Mask Attributes & mask    & mask          & mask     & mask          \\ \hline
	\end{tabular}
\caption{Implementation details for general augmentation. Edge refers all edges that reach out from the corresponding node. - denotes no change.}\label{app-aug}
\end{table}
\vspace{-2em}

\begin{figure}[!h]
	\centering
	\begin{subfigure}[b]{0.23\textwidth}
		\includegraphics[width=\textwidth, trim=0 6cm 0 6cm, clip]{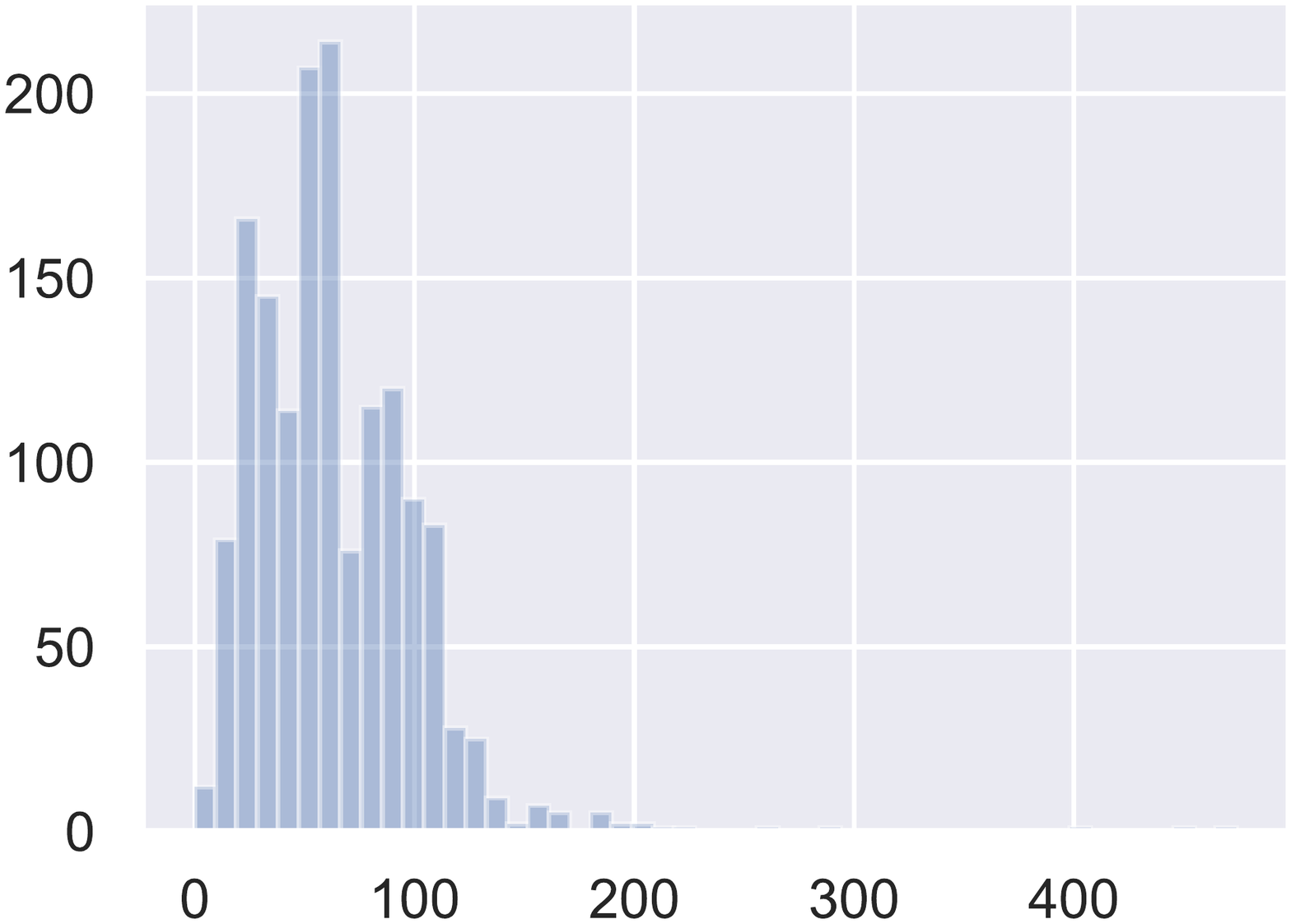}
		\caption{Function group rule}
		\label{app-1}
	\end{subfigure}
	\begin{subfigure}[b]{0.23\textwidth}
		\includegraphics[width=\textwidth, trim=0 6cm 0 6cm, clip]{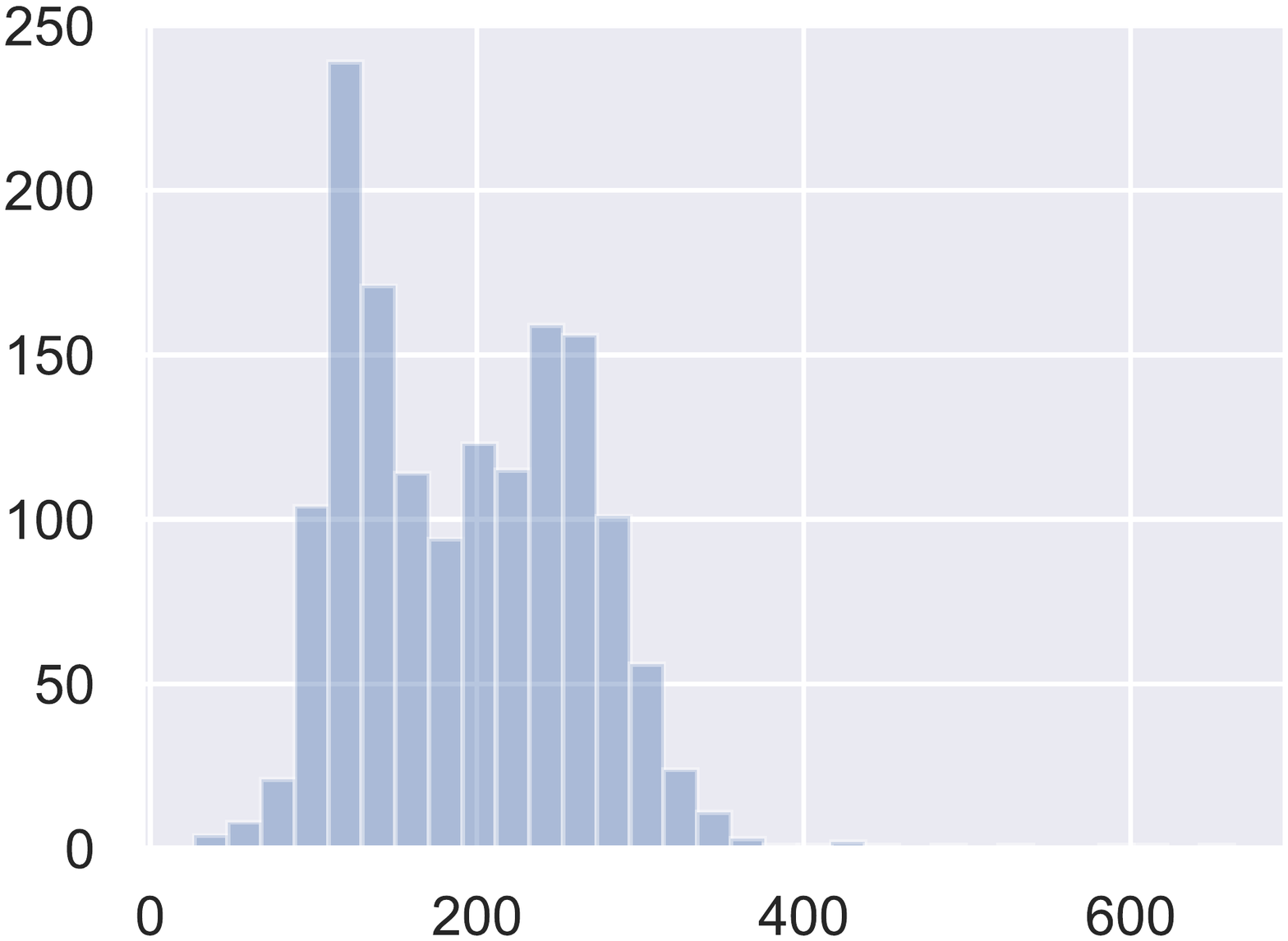}
		\caption{General carbon rule}
		\label{app-2}
	\end{subfigure}
	\caption{Distribution of augmentations that can be generated by proposed augmentation rules (dataset: bace).}\label{app-dist}
\end{figure}

\noindent Figure \ref{app-dist} shows the distribution of number of augmentations that can be generated by applying MoCL-DK1 (left: from rules of substituting functional groups; right: from rules of adding/dropping general carbons). Other datasets reveal the same pattern therefore we do not include them due to space limit. We see that MoCL-DK1 can generate considerable number of augmentations for the molecules. If we apply MoCL-DK multiple times (MoCL-DK3, MoCL-DK5), the number of possible products can further increase drastically.

\newpage
\subsection*{Proof of Lemma~\ref{lemma1}}
	Assume the projection head $g$ is an identity mapping, i.e., $\mathbf{z} = g(\mathbf{h})=\mathbf{h}$, and the similarity function $s(\cdot, \cdot)$ is inner product, i.e., $s(\mathbf{z}_i, \mathbf{z}_j) = \mathbf{z}_i^T \mathbf{z}_j$. Consider 1-nearest neighbor of each graph in the batch for global structure information, and $\lambda=1$, the objective $\mathcal{L}_i$ is equivalent to the following:
	\begin{align*}
	\mathcal{L}_i \propto& \sum_{j\ne i} \underbrace{\|\mathbf{z}^1_i-\mathbf{z}^2_i\|^2 - \|\mathbf{z}^1_i-\mathbf{z}^2_j\|^2}_\text{local contrast view 1} + \underbrace{\|\mathbf{z}^2_i-\mathbf{z}^1_i\|^2 - \|\mathbf{z}^2_i-\mathbf{z}^1_j\|^2}_\text{local contrast view 2} \nonumber\\
	& + \sum_{j\ne k, k\in \mathcal{N}_i} \underbrace{\|\mathbf{z}_i-\mathbf{z}_k\|^2 - \|\mathbf{z}_i-\mathbf{z}_j\|^2}_\text{global contrast} + Const.
	\end{align*} 

\begin{proof}
\begin{align*}
\mathcal{L}_i &= \log \frac{\sum_{j\ne i}^{n}e^{s(\mathbf{z}^1_i, \mathbf{z}^2_j)}}{e^{s(\mathbf{z}^1_i, \mathbf{z}^2_i)/\tau}} + \log \frac{\sum_{j\ne i}^{n}e^{s(\mathbf{z}^2_i, \mathbf{z}^1_j)/\tau}}{e^{s(\mathbf{z}^2_i, \mathbf{z}^1_i)/\tau}} \nonumber\\ 
& \quad + \log \frac{\sum_{j\ne k, k\in \mathcal{N}_i}^{n}e^{s(\mathbf{z}_i, \mathbf{z}_j)/\tau}}{e^{s(\mathbf{z}_i, \mathbf{z}_k)}} \\
&=\log \sum_{j\ne i}^{n}e^{s(\mathbf{z}_i^1, \mathbf{z}_j^2)/\tau - s(\mathbf{z}_i^1, \mathbf{z}_i^2)/\tau} + \log \sum_{j\ne i}^{n}e^{s(\mathbf{z}_i^2, \mathbf{z}_j^1)/\tau - s(\mathbf{z}_i^2, \mathbf{z}_i^1)/\tau} \nonumber \\
& \quad + \log \sum_{j\ne k, k\in \mathcal{N}_i}^{n}e^{s(\mathbf{z}_i, \mathbf{z}_j)/\tau - s(\mathbf{z}_i, \mathbf{z}_k)/\tau}
\end{align*}
By applying first-order Taylor expansion we have:

\begin{align*}
\mathcal{L}_i &\approx \sum_{j\ne i}^{n}e^{s(\mathbf{z}_i^1, \mathbf{z}_j^2)/\tau - s(\mathbf{z}_i^1, \mathbf{z}_i^2)/\tau} + \sum_{j\ne i}^{n}e^{s(\mathbf{z}_i^2, \mathbf{z}_j^1)/\tau - s(\mathbf{z}_i^2, \mathbf{z}_i^1)/\tau} \nonumber \\
& \quad + \sum_{j\ne k, k\in \mathcal{N}_i}^{n}e^{s(\mathbf{z}_i, \mathbf{z}_j)/\tau - s(\mathbf{z}_i, \mathbf{z}_k)/\tau} - 3\\
& \approx \frac{1}{\tau} \big[\sum_{j\ne i}^{n} {s(\mathbf{z}_i^1, \mathbf{z}_j^2) - s(\mathbf{z}_i^1, \mathbf{z}_i^2)} + \sum_{j\ne i}^{n} {s(\mathbf{z}_i^2, \mathbf{z}_j^1) - s(\mathbf{z}_i^2, \mathbf{z}_i^1)} \nonumber\\
& \quad + \sum_{j\ne k, k\in \mathcal{N}_i}^{n} {s(\mathbf{z}_i, \mathbf{z}_j) - s(\mathbf{z}_i, \mathbf{z}_k)}\big] - 3 \\
& = \frac{1}{\tau} \big[\sum_{j\ne i}^{n} {\mathbf{z}_i^{1_{T}} \mathbf{z}_j^2 - \mathbf{z}_i^{1_{T}} \mathbf{z}_i^2} + \sum_{j\ne i}^{n} {\mathbf{z}_i^{2_{T}} \mathbf{z}_j^1 - \mathbf{z}_i^{2_{T}} \mathbf{z}_i^1} \nonumber\\
& \quad + \sum_{j\ne k, k\in \mathcal{N}_i}^{n} {\mathbf{z}_i^T \mathbf{z}_j - \mathbf{z}_i^T \mathbf{z}_k}\big] - 3\\
& = \frac{1}{2\tau} \big[\sum_{j\ne i}^{n} \|\mathbf{z}^1_i-\mathbf{z}^2_i\|^2 - \|\mathbf{z}^1_i-\mathbf{z}^2_j\|^2+\|\mathbf{z}^2_i-\mathbf{z}^1_i\|^2 - \|\mathbf{z}^2_i-\mathbf{z}^1_j\|^2 \nonumber\\
& \quad + \sum_{j\ne k, k\in \mathcal{N}_i}^{n} \|\mathbf{z}_i-\mathbf{z}_k\|^2 - \|\mathbf{z}_i-\mathbf{z}_j\|^2\big] - 3\\
& \propto \sum_{j\ne i} \|\mathbf{z}^1_i-\mathbf{z}^2_i\|^2 - \|\mathbf{z}^1_i-\mathbf{z}^2_j\|^2+\|\mathbf{z}^2_i-\mathbf{z}^1_i\|^2 - \|\mathbf{z}^2_i-\mathbf{z}^1_j\|^2 \nonumber\\
& \quad + \sum_{j\ne k, k\in \mathcal{N}_i} \|\mathbf{z}_i-\mathbf{z}_k\|^2 - \|\mathbf{z}_i-\mathbf{z}_j\|^2 -6\tau
\end{align*} 
\end{proof}

\end{document}

%% file: intro.tex
% !TEX root = ./main.tex

% graph neural network
Graph neural networks (GNNs) has been demonstrated to achieve state-of-the-art performance on graph-related tasks such as node classification \cite{kipf2016semi, velivckovic2017graph, wu2019simplifying}, link prediction \cite{zitnik2018modeling} and graph classification \cite{velivckovic2017graph, gilmer2017neural, xu2018powerful}. It has also been frequently used in the biomedical domain recently to tackle drug-related problems \cite{stokes2020deep, sakai2021prediction, mercado2020graph}. However, like most deep learning architectures, it requires large amount of labeled data to train whereas task-specific labels in real world are often of limited size (e.g., in biomedical domain, requiring labels such as drug responses from biological experiments is always expensive and time consuming). Therefore, pretraining schemes on GNNs have been actively explored recently. 

% GNN pretraining and contrastive learning on graphs
One line of works focuses on designing pretext tasks to learn node or graph representations without labels. The predefined tasks include graph reconstruction~\cite{kipf2016variational, hu2020gpt, you2020does} and context prediction~\cite{peng2020self, hu2019strategies}. The other line follows a contrastive learning framework from computer vision domain~\cite{chen2020simple, wu2018unsupervised}, in which two augmentations are generated for each data and then fed into an encoder and a projection head. By maximizing the mutual information between the two augmented views, the model is able to learn representations that are invariant to transformations. In particular, \cite{you2020graph} proposed four types of augmentations for general graphs and demonstrated that contrastive learning on graphs is able to produce representations that are beneficial for downstream tasks.  

% challenges of current graph CL
However, contrastive learning on graphs has its unique challenges. First, the structural information and semantics of the graphs varies significantly across domains (e.g., social network v.s. molecular graphs), thus it is difficult to design universal augmentation scheme that fits all scenarios. It has been shown that general augmentations can be harmful under a specific domain context \cite{you2020graph}. Second, most current graph contrastive learning frameworks learn invariant representations while neglect the global structure of the entire data \cite{asano2019self}, e.g., some graphs should be closer in the embedding space due to their structural similarity. Nevertheless, modeling similarity between graphs itself is still a difficult problem \cite{bai2019unsupervised}. Third, the contrast schemes are not unique because graph tasks can happen at different levels, e.g., node-graph contrast \cite{hassani2020contrastive}, node-node contrast \cite{zhu2020graph}, graph-graph contrast \cite{you2020graph} are all possible contrast schemes.

Besides these unique challenges for graphs, contrastive learning itself also has unsolved problems. For example, accurately estimating mutual information in high dimension is difficult \cite{poole2019variational}. The connection between mutual information maximization and the success of contrastive learning is still not clear. In fact, \cite{tschannen2019mutual} found the connection is actually weak, while instead metric learning shares some intrinsic connections with contrastive learning. These findings also motivate us to pay more attention to the role of augmentation schemes and global semantics of the data in order to improve contrastive learning on graphs.

% domain knowledge with machine learning  
Therefore, in this paper, we aim to tackle the aforementioned challenges in the context of biomedical domain, where molecular graphs are present. 
Our goal is to improve representations by infusing domain knowledge into the augmentation and constrast schemes. 
We propose to leverage both local-level and global-level domain knowledge to assist contrastive learning on molecular graphs. 
In particular, unlike general augmentations in which nodes and edges in a graph are randomly perturbed, we propose a new augmentation scheme called substructure substitution where a valid substructure in a molecule is replaced by a bioisostere that introduces variation without altering the molecular properties too much. 
The substitution rules are derived from domain resource and we regard it as local-level domain knowledge. 
The global-level domain knowledge encodes the global similarities between graphs. 
We propose to utilize such information to learn richer representations via a double contrast objective.

Leveraging domain knowledge to assist contrastive learning has rarely been explored in literature and our work is the first to make this attempt. In summary, our contributions are as follows:

\begin{itemize}
	\item We propose a new augmentation scheme for molecular graphs based on local-level domain knowledge such that the semantics of graphs do not change in the augmentation process.
	\item We propose to encode global structure of the data into graph representations by adding a global contrast loss utilizing the similarity information between molecular graphs.
	\item We provide theoretical justifications that the learning objective is connected with triplet loss in metric learning which shed light on the effectiveness of the entire framework.
	\item We evaluate MoCL on various molecular datasets under both linear and semi-supervised settings and demonstrate its superiority over the state-of-the-art methods.
\end{itemize}

%The rest of the paper is organized as follows: we summarize related work in Section 2; the proposed framework is presented in Section 3; experimental results are shown in Section 4 and conclusion reaches at Section 5.

%% file: related_work.tex
% !TEX root = ./main.tex
% MoCL
%Our work is related to self-supervised learning on graphs, graph contrastive learning and pretraining graph neural networks. Below we summarize the related work from these directions.

%evaluation protocol not finshed, put it in exp

\noindent \textbf{Self-supervised learning on graphs.} A common strategy for learning node (graph) representation in an unsupervised manner is to design pretext tasks on unlabled data. For node-level tasks, You et al. \cite{you2020does} proposed three types of self-supervised tasks: node clustering, graph partition and graph completion to learn node representations. Peng et al. \cite{peng2020self} proposed to predict the contextual position of a node relative to the other to encode the global topology into node representations. GPT-GNN \cite{hu2020gpt} designed generative task in which node attributes and edges are alternatively generated such that the likelihood of a graph is maximized. After that, the pretrained GNN can be used for any downstream tasks. For graph level tasks, Hu et al. \cite{hu2019strategies} first designed two tasks, predicting neighborhood context and node attributes to learn meaningful node representations, then using graph-level multi-task pretraining to refine the graph representation. GROVER \cite{rong2020self} incorporated GNN into a Transformer-style architecture and learned node embedding by predicting contextual property and motif labels. Other works \cite{shang2019pre, yasunaga2020graph, sun2020multi} utilized similar strategies for either node or graph level pretraining in the context of a more specific task or domain.

\noindent \textbf{Contrastive learning on graphs.} Contrastive learning on graphs can be categorized into two groups. One group aims to encode structure information by contrasting local and global representations. For example, DGI \cite{velickovic2019deep} proposed to maximize the mutual information between node embedding and graph summary vector to learn node representations that capture the graph semantics. InfoGraph \cite{sun2019infograph} extended DGI to learn graph-level representations and further proposed a variant for semi-supervised scenarios. Another group aims to learn representations that are invariant to transformations, following the idea of contrastive learning on visual representations \cite{chen2020simple, wu2018unsupervised, dosovitskiy2014discriminative}, where two augmentations (views) of an image are generated and fed into an encoder and a projection head, after which their mutual information is maximized. Similarly, You et al. \cite{you2020graph} explored four types of augmentations for general graphs and demonstrated that the learned representations can help downstream tasks. Instead of general corruption, \cite{hassani2020contrastive} used graph diffusion to generate the second view and performed contrast between node and graph from two views. GCA \cite{zhu2020graph} proposed adaptive augmentation such that only unimportant nodes and edges are perturbed. However, GCA is focused on network data and not suitable for molecular graphs. Instead of focusing on augmentation views, MICRO-Graph \cite{zhang2020motif} proposed to contrast based on sub-graphs (motifs). GCC \cite{qiu2020gcc} proposed to use random walk to generate subgraphs and contrast between them. 

\noindent \textbf{Evaluation protocols.} There exist various evaluation schemes for graph level self-supervised learning. Most prior works \cite{sun2019infograph, hu2019strategies, you2020graph, zhang2020motif} adopt the linear evaluation protocol where a linear classifier is trained on top of the representations. \cite{sun2019infograph, you2020graph, zhang2020motif} also adopt the semi-supervised protocol where only a small fraction of labels are available for downstream tasks. Other works \cite{hu2019strategies, you2020graph, rong2020self} also explore the transfer learning setting in which the pretrained model is applied to other datasets.

 %Therefore, what augmentations to generate and how to contrast are the
 %Since our work focus on graph-level representation learning, and there are various evaluation schemes for graph self-supervised learning, we summarize the evaluation protocols that related works use. 

%% file: method.tex
% !TEX root = ./main.tex

 \begin{figure}[tb]
	\centering
	\includegraphics[width=0.47\textwidth, trim=0.0cm 6cm 0.0cm 7cm, clip]{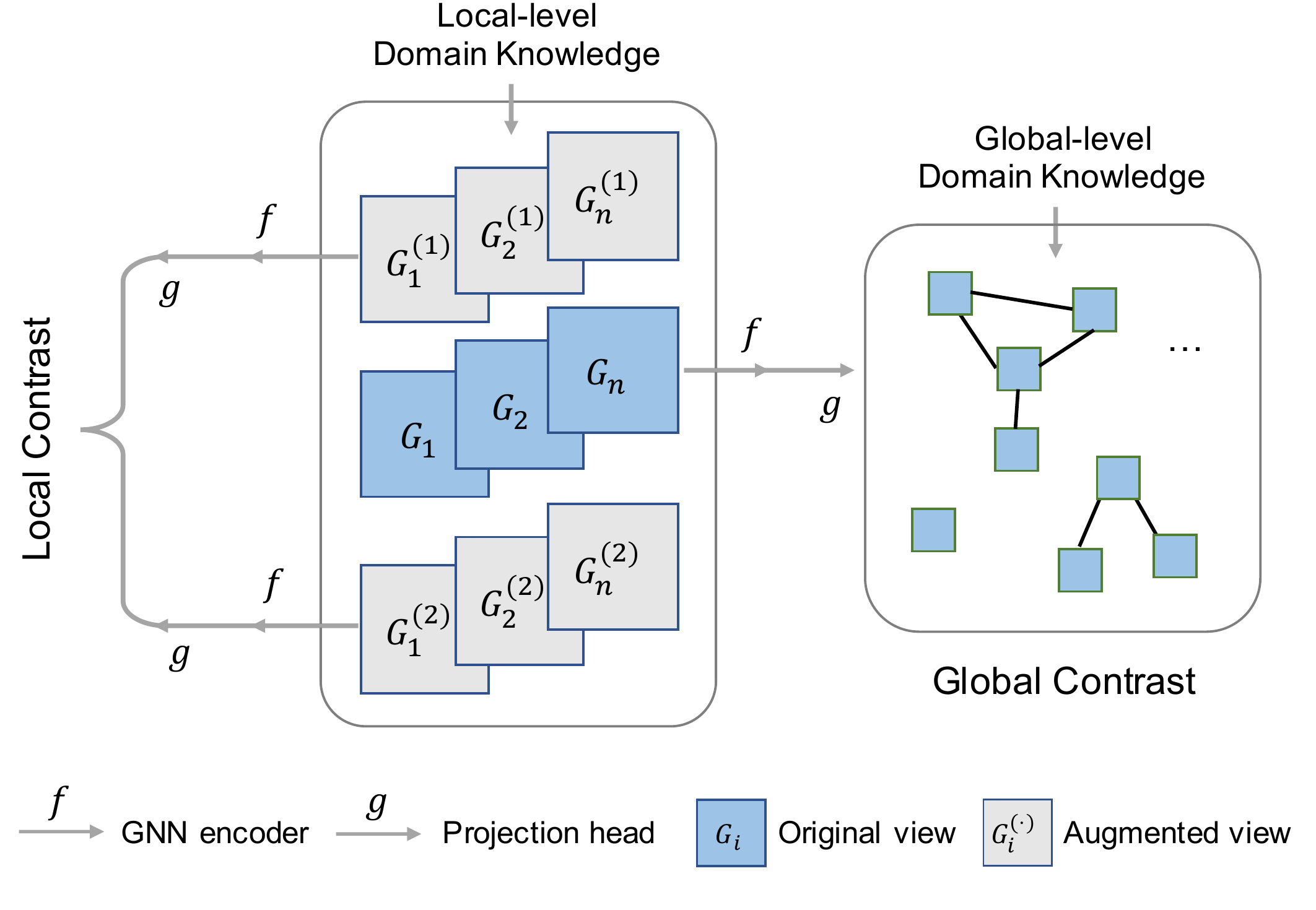}
	\vspace{-0.5em}
	\caption{Overall framework of MoCL. First, two augmented views are generated from local-level domain knowledge. Then, together with the original view (blue), they are fed into the GNN encoder and projection head. The local-level contrast maximizes the mutual information (MI) between two augmented views. The global-level contrast maximizes the MI between two similar graphs, where the similarity information is derived from global-level domain knowledge.}\label{fw}

\end{figure}

\subsection{Problem Definition}
A (molecular) graph can be represented as $\mathcal{G=(V, E)}$, where $\mathcal{V}=\{v_1, v_2,.., v_{|V|}\}$ and $\mathcal{E = V \times V}$ denotes node and edge set respectively. Let $\mathbf{X} \in \mathbb{R}^{|V|\times d_1}$ be the feature matrix for all nodes in a graph, $\mathbf{A} \in \mathbb{R}^{|V| \times |V|}$ the adjacency matrix and $\mathbf{E} \in \mathbb{R}^{|\mathcal{E}| \times d_2}$ the edge features, our goal is to learn a graph encoder $\mathbf{h} = f(\mathbf{X, A, E}) \in \mathbb{R}^{d'}$ which maps an input graph to a vector representation without the presence of any labels. The learned encoder and representations can be used for downstream tasks directly or via finetune. %A good pretrained encoder is the one that benefits downstream tasks as compared to train from scratch.      

\subsection{Contrastive Learning Framework}
%(e.g., a graph neural network),
% (e.g., multi-layer perceptron)
In a conventional contrastive learning framework (Fig. \ref{fw} left), for each graph $G_i$, two augmentation operators $t_1$ and $t_2$ are sampled from the family of all operators $\mathcal{T}$, and applied to $G_i$ to obtain two correlated views $G^{1}_i=t_1(G_i)$ and $G^{2}_i=t_2(G_i)$. We use numbers in the \textbf{superscript} to represent different \textbf{views} throughout the paper. The correlated views are fed into a graph encoder $f$, producing graph representations $\mathbf{h}_i^{1}$ and $\mathbf{h}_i^{2}$, which are then mapped into an embedding space by a projection head $g$, yielding $\mathbf{z}_i^{1}$ and $\mathbf{z}_i^{2}$. The goal is to maximize the mutual information between the two correlated views in the embedding space via Eq (\ref{loss}).

\vspace{-1em}
\begin{align}
\mathcal{L}^{\text{local}} = \frac{1}{n} \sum\nolimits_{i=1}^{n} \mathcal{L}_i^{\text{local}}, \label{loss}
\end{align}
and the loss for each sample $\mathcal{L}_i^{\text{local}}$ can be written as:

\vspace{-1em}
\begin{align}\hspace{-0.08in}
\mathcal{L}_i^{\text{local}} &= \mathcal{L}^1_i + \mathcal{L}^2_i \nonumber \\ \label{loss-i}
& = -\log \frac{e^{s(\mathbf{z}_i^{1}, \mathbf{z}_i^{2})/\tau}}{\underbrace{\sum\nolimits_{j=1, j\ne i}^{n} e^{s(\mathbf{z}_i^{1}, \mathbf{z}_j^{2})/\tau}}_\text{view 1 contrasts view 2}} -\log \frac{e^{s(\mathbf{z}_i^{2}, \mathbf{z}_i^{1})/\tau}}{\underbrace{\sum\nolimits_{j=1, j\ne i}^{n} e^{s(\mathbf{z}_i^{2}, \mathbf{z}_j^{1})/\tau}}_\text{view 2 contrasts view 1}},
\end{align}
%We use graph neural network for $f$ and 2-layer perceptron for $g$. 
where $n$ is the batch size, $s(\cdot,\cdot)$ is a function which measures the similarity of the two embeddings, $\tau$ is a scale parameter. The two correlated views $\mathbf{z}^1_i$ and $\mathbf{z}^2_i$ are regarded as positive pair while the rest pairs in the batch are regarded as negative pairs. The objective aims to increase the probability of occurrences of positive pairs as opposed to negative ones. Note that the negative pairs can be formed in two directions. If $\mathbf{z}^1_i$ is the anchor, all $\mathbf{z}^2_j$ in view 2 are contrasted; if $\mathbf{z}^2_i$ is the anchor, all $\mathbf{z}^1_j$ in view 1 are contrasted. Thus the loss for each sample consists of two parts as showed in Eq (\ref{loss-i}).

% necessity of projection head g
% sensitivity analysis

 \begin{figure}[tb]
	\centering
	\includegraphics[angle=270, width=0.49\textwidth, trim=2cm 0.0cm 2cm 0cm, clip]{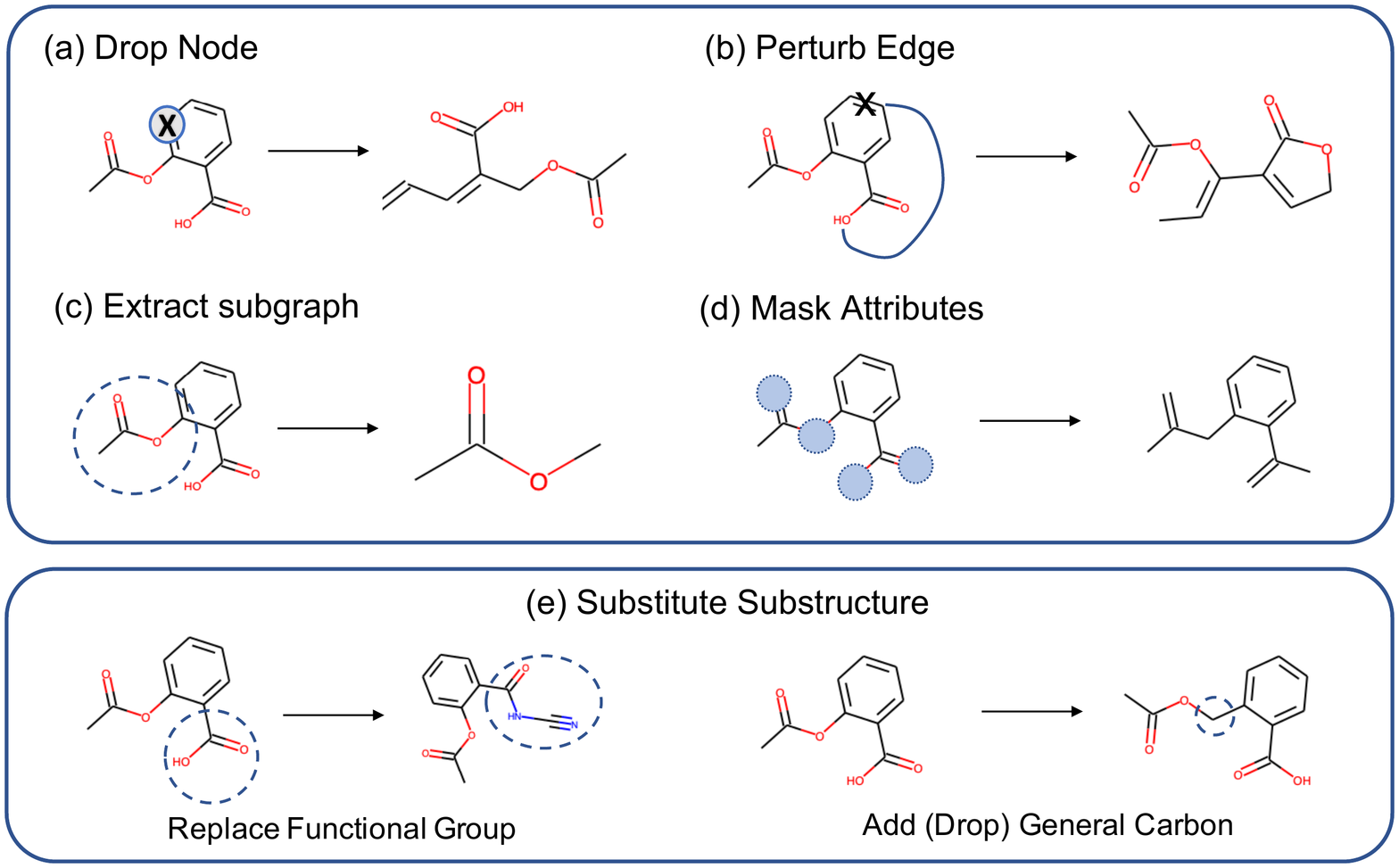}
	\vspace{-0.5em}
	\caption{Augmentation comparison. Upper: conventional augmentations that may alter the graph semantics. Lower: proposed augmentation in which valid substructures are replaced by bioisosteres that share similar properties.}\label{aug}
\end{figure}

\subsection{Local-level Domain Knowledge}
Most existing approaches adopt random corruption during augmentation. For example, \cite{zhu2020graph} proposed four types of augmentations for general graphs (Fig. \ref{aug} upper). However, such random corruption may alter the semantics of molecular graphs. For node dropping and edge perturbation, the resulting molecule is rarely biologically proper, e.g., dropping a carbon atom in the phenyl ring of aspirin breaks the aromatic system and results in an alkene chain (Fig. \ref{aug}a); perturbing the connection of aspirin might introduce a five-membered lactone (Fig. \ref{aug}b), which may drastically change the molecular properties. For subgraph extraction, the resulting structure is arbitrary and not representative for molecular functionality, e.g., methyl acetate is a sub group of aspirin (Fig. \ref{aug}c), but also frequently shown in other compounds such as digitoxin and vitamin C with diverse chemical structures and biological effects. Enforcing high mutual information between such augmentation pairs may produce suboptimal representations for downstream tasks. This phenomenon has also been observed in \cite{zhu2020graph} that edge perturbation deteriorates the performance of certain molecular tasks. Among the general augmentations, only attribute masking (Fig. \ref{aug}d) does not violate the biological assumptions since it does not change the molecule, it only masks part of the atom and edge attributes.

Therefore, we aim to infuse domain knowledge to assist the augmentation process. We propose a new augmentation operator called \emph{substructure substitution}, in which a valid substructure in a molecule is replaced by a bioisostere \cite{meanwell2011synopsis} which produces a new molecule with similar physical or chemical properties as the original one (Fig. \ref{aug}e). We compile 218 such rules from domain resource \footnote{https://www.schrodinger.com/drug-discovery}. Each rule consists of a source substructure and a target substructure represented by SMARTS string \footnote{https://www.daylight.com/dayhtml/doc/theory/theory.smarts.html}. A sample rule is as follows:

\begin{verbatim}
  [#6:2][#6:1](=O)[O;-,H1] >> [*:2][c:1]1nn[nH]n1
\end{verbatim}
indicating the transition from left substructure (carboxylic acid) to the right one (nitrogen heterocycle). The substitution rules have 36 unique source substructures which can be categorized into 8 groups. We summarize the statistics of the rules in Table \ref{rules}. Note that target substructures are all unique and different. The original 218 substitution rules mostly happen at molecular positions where heteroatoms (heavy atoms that are not C or H) and aromatic rings are presented, therefore the variation for general carbon groups is limited. Under the common assumption that changing a few general carbon atoms will not alter the molecular property too much, we add 12 additional rules to subtract and add general carbon groups from and to a molecule. Some sample rules are:
\begin{verbatim}
   [*:1][CH2][CH2][*:2] >> [*:1][*:2]  (drop)
   [*:1]-[*:2] >> [*:1]CC[*:2]         (add)
\end{verbatim}
Thus, MoCL consists of 230 rules in total to generate molecule variants that share similar properties. All the rules and code are available at https://github.com/illidanlab/MoCL-DK.

\begin{table}[tb]
	\begin{tabular}{cccc}
		\hline
		Group   & \# source & \# target & Formula \\ \hline
		CA      & 1             & 68            & RCOO       \\
		Ester   & 1             & 7             & RCOOR'		    \\
		Ketone  & 1             & 15            & ROR'       \\
		Phenyl  & 22            & 36            & Aromatic Rings   \\
		Tbutyl  & 1             & 10            & C4       \\
		dsAmide & 4             & 18            & RONR'R''       \\
		msAmide & 2             & 32            & RONR'       \\
		nsAmide & 4             & 32            & RON       \\ \hline
		Total   & 36            & 218           & -       \\ \hline
	\end{tabular}
	\caption{Source and target statistics for substitution rules. R/R'/R'' represent  arbitrary carbon-containing groups.}\label{rules}
\end{table}

Moreover, since the source substructures in the rules are very common, a molecule may contain multiple source substructures or multiple copies of the same substructure in the rule, the proposed augmentation can be applied multiple times to generate variants with much more diversity. A notable difference between proposed augmentation and general augmentation is that the proposed rules are not guaranteed to be applicable to a molecule after it changes, therefore when applying proposed augmentation multiple times, we need to update the rule availability accordingly at each round. We summary the proposed augmentation procedure in Alg. \ref{alg-1}.

\subsection{Global-level Domain Knowledge}
Maximizing mutual information between correlated views learns transformation-invariant representations. However, it may neglect the global semantics of the data. For example, some graphs should be closer in the embedding space since they share similar graph structures or semantics from domain knowledge. For molecular graphs, such information can be derived from multiple sources. For general graph structure, extended connectivity fingerprints (ECFPs) \cite{rogers2010extended} encode the presence of substructures for molecules and are widely used to measure the structural similarity between molecular graphs. Drug-target networks~\cite{Ramsundar-et-al-2019} record the drug-protein interaction information which is one of the most informative biological activity measures. In this section, we first define graph similarity from general molecular graphs, then we propose two ways to incorporate the global semantics into our learning framework.

\subsubsection{Similarity calculation}
Given the ECFP of two molecules, $e_1, e_2$ $\in \{0,1\}^m$ where $m$ is the vector length and $1$ indicates the presence of certain substructures, the similarity of $e_1$ and $e_2$ can be calculated as the Tanimoto coefficient \cite{bajusz2015tanimoto}:

\vspace{-0.5em}
\begin{align}
	s(e_1, e_2) = \frac{N_{12}}{N_1+N_2-N_{12}}, \label{sim}
\end{align}
where $N_1, N_2$ denotes the number of $1$s in $e_1, e_2$ respectively, and $N_{12}$ denotes the number of $1$s in the intersection of $e_1, e_2$. The resulted coefficient $s(e_1, e_2) \in [0, 1]$ and a larger value indicates higher structural similarity. Similarly, for drug-target network, $e_1, e_2$ $\in \{0,1\}^m$ becomes the interaction profile of a drug to all proteins where $m$ is the total number of proteins. The drug similarity can be calculated the same as Eq.~(\ref{sim}).

\begin{algorithm}[tb]
	\DontPrintSemicolon % Some LaTeX compilers require you to use \dontprintsemicolon instead
	\KwIn{Molecule graph $G$, repeat time $R$, rules $\mathcal{T}$}
	\KwOut{Augmented graph $G'$}
	\For{$r = 1$ \textbf{to} $R$} {
		\While{$\mathcal{T}$}{
			sample $t \sim \mathcal{T}$ \quad \quad \quad \text{\# one augmentation rule}\\
			$\{G^1, G^2,..,G^k\} = t(G)$ \quad \text{\# all possible products}\\
			random choose $G=G^i$ \\
			update available $\mathcal{T}$ \text{\# rules may no longer be valid} \\
			break; 
		}
	}
	$G'=G$\\
	\Return{$G'$}\;
	\caption{Pseudocode of domain augmentation.}
	\label{alg-1}
\end{algorithm}

\subsubsection{Global-level Objective}

We propose two strategies for using the global similarity information. One strategy is to use it as direct supervision. Given embeddings of two original graphs $\mathbf{z}_i$ and $\mathbf{z}_j$, we measure the similarity between them as $\theta(\mathbf{z}_i, \mathbf{z}_j)=\frac{\mathbf{z}_i^{T}\mathbf{z}_j}{\|\mathbf{z}_i\|\|\mathbf{z}_j\|}$. We optimize the similarity using least square loss as follows:
\begin{align*}
\mathcal{L}^{\text{global}}_i = \sum\nolimits_{j \ne i}\mathcal{L}^{\text{global}}_{ij} =\sum\nolimits_{j \ne i} \|\theta(\mathbf{z}_i, \mathbf{z}_j)-s_{i,j}\|_2^2,
\end{align*}
where $s_{i, j}$ is the similarity from Eq.~(\ref{sim}).

The second strategy is to utilize a contrastive objective in which similar graph pairs have higher mutual information as compared to the background. The objective is written as:
\begin{align*}
\mathcal{L}^{\text{global}}_i =  -\log \frac{\sum_{j=1, j\in \mathcal{N}_i}^{n} e^{s(\mathbf{z}_i, \mathbf{z}_j)/\tau}}{\sum_{j=1, j\notin \mathcal{N}_i}^{n} e^{s(\mathbf{z}_i, \mathbf{z}_j)/\tau}},
\end{align*}
where $\mathcal{N}_i$ refers the neighbors of graph $i$. The neighbors can be derived from global similarity by setting a threshold or a neighborhood size. The global loss for all graphs thus becomes:
\begin{align}
\mathcal{L}^{\text{global}} = \frac{1}{n}\sum\nolimits_{i=1}^{n}\mathcal{L}_i^{\text{global}}. \label{loss-g}
\end{align}
Finally, the full objective of the proposed MoCL can be written as:
\begin{align}
\mathcal{L} = \mathcal{L}^{\text{local}} + \lambda \mathcal{L}^{\text{global}}, \label{loss-all}
\end{align}
where $\lambda$ is a tuning parameter that controls the emphasis between local loss and global loss. We summarize the pseudo code of the entire framework in Alg. \ref{alg-2}.

\begin{algorithm}[tb]
	\DontPrintSemicolon % Some LaTeX compilers require you to use \dontprintsemicolon instead
	\KwIn{Molecule graphs $G$, rules $\mathcal{T}$, hyper parameter $\lambda$, number of epochs $M$}
	\KwOut{Graph encoder $f$}
	\For{$m = 1$ \textbf{to} $M$} {
		\For{$iter=1$ \textbf{to} $\text{max\_iter}$}{
		$G^1=\text{Alg.}1(G, \mathcal{T}), G^2=\text{Alg}.1(G, \mathcal{T})$\\
		$\mathbf{h}^1=f(G^1), \mathbf{h}^2=f(G^2),  \mathbf{h}=f(G)$\\
		$\mathbf{z}^1=g(\mathbf{h}^1), \mathbf{z}^2=g(\mathbf{h}^2), \mathbf{z}=g(\mathbf{h})$\\
		Calculate local loss by Eq. (\ref{loss})\\
		Calculate global loss by Eq. (\ref{loss-g})\\
		Optimize $f$ and $g$ using Eq. (\ref{loss-all})
	}
	}
	\Return{$f$}
	\caption{Pseudocode of proposed framework.}
	\label{alg-2}
\end{algorithm}

\subsection{Connection to Metric Learning}
It has been well studied that optimizing objective Eq.~(\ref{loss}) is equivalent to maximizing a lower bound of the mutual information between the correlated views, also a lower bound of the mutual information between input and the hidden representations \cite{oord2018representation, cover1999elements}. Formally, denote $\mathbf{Z}^1$ and $\mathbf{Z}^2$ as the random variables for the embeddings of augmentations, $\mathbf{X}$ the variable for original input features:
\begin{align*}
\mathcal{L}^\text{local} \le I(\mathbf{Z}^1; \mathbf{Z}^2) \le I(\mathbf{X}; \mathbf{Z}^1, \mathbf{Z}^2). 
\end{align*}

Beyond mutual information maximization, in this section, we provide additional justification for the proposed method from the perspective of metric learning, which unifies the local and global objectives. We show the following important result:
\begin{lemma}\label{lemma1}
	Assume the projection head $g$ is an identity mapping, i.e., $\mathbf{z} = g(\mathbf{h})=\mathbf{h}$, and the similarity function $s(\cdot, \cdot)$ is inner product, i.e., $s(\mathbf{z}_i, \mathbf{z}_j) = \mathbf{z}_i^T \mathbf{z}_j$. Consider 1-nearest neighbor of each graph in the batch for global structure information, and $\lambda=1$, the objective $\mathcal{L}_i$ is equivalent to the following:
	\begin{align*}
	\mathcal{L}_i \propto& \sum_{j\ne i} \underbrace{\|\mathbf{z}^1_i-\mathbf{z}^2_i\|^2 - \|\mathbf{z}^1_i-\mathbf{z}^2_j\|^2}_\text{local contrast view 1} + \underbrace{\|\mathbf{z}^2_i-\mathbf{z}^1_i\|^2 - \|\mathbf{z}^2_i-\mathbf{z}^1_j\|^2}_\text{local contrast view 2} \nonumber\\
	& + \sum_{j\ne k, k\in \mathcal{N}_i} \underbrace{\|\mathbf{z}_i-\mathbf{z}_k\|^2 - \|\mathbf{z}_i-\mathbf{z}_j\|^2}_\text{global contrast} + Const.
	\end{align*} 
\end{lemma}
The lemma above connects the objective design to the metric learning. The equation consists of three triplet losses \cite{chechik2009large} which corresponds to the two local losses and the global loss respectively. As such, the MoCL objective aims to pull close the positive pairs while pushing away the negative pairs from both local and global perspective. Detailed proofs can be found in Appendix.

%% file: exp.tex
% !TEX root = ./main.tex

% question to ask
% synthetic experiment
% real daat experiment
% dataset 
% GNN arch and g, number of neurons
% baseline
% experiment settings 
% abalation study
In this section, we conduct extensive experiments to demonstrate the proposed method by answering the following questions:

$\mathbf{Q1}.$ Does local-level domain knowledge (MoCL-DK) learns better representations than general augmentations? How does combination of different augmentations behave?

$\mathbf{Q2}.$ Does global-level domain knowledge (MoCL-DK-G) further improve the learned representations? Do the two proposed global losses perform the same?

$\mathbf{Q3}.$ How do the hyper-parameters ($\lambda$, neighbor size) involved in MoCL affect the model performance?

\begin{table}[tb]
	\begin{tabular}{cccccc}
		\hline 
		Dataset & \# Tasks & Size  & Avg. Node & Avg. Degree\\ \hline
		bace    & 1       & 1513  & 34.1      & 36.9       \\
		bbbp    & 1       & 2050  & 23.9      & 25.8       \\
		clintox & 2       & 1483  & 26.1      & 27.8       \\
		mutag   & 1       & 188   & 17.8      & 19.6       \\
		sider   & 27      & 1427  & 33.6      & 35.4       \\
		tox21   & 12      & 7831  & 18.6      & 19.3       \\
		toxcast & 617     & 8597  & 18.7      & 19.2       \\ \hline
	\end{tabular}
	\caption{Basic statistics for all datasets}\label{exp-stats}
	\vspace{-0.3in}
\end{table}

\begin{figure*}[tb]
	\centering
	\begin{subfigure}[b]{0.32\textwidth}
		\includegraphics[width=\textwidth, trim=0 5.5cm 0 5cm, clip]{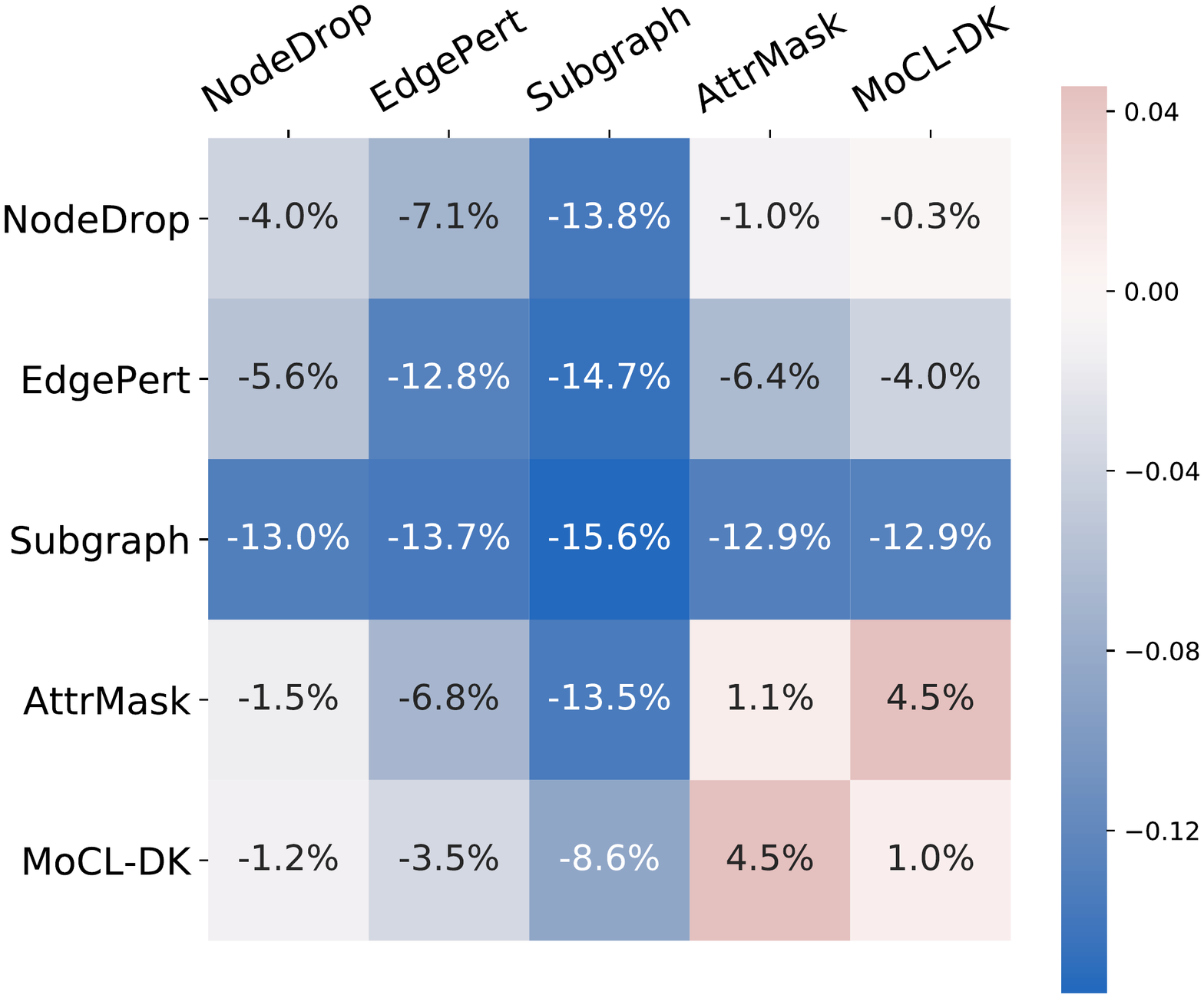}
		\caption{bace}
		\label{d-1}
	\end{subfigure}
	\begin{subfigure}[b]{0.32\textwidth}
		\includegraphics[width=\textwidth, trim=0 5.5cm 0 5cm, clip]{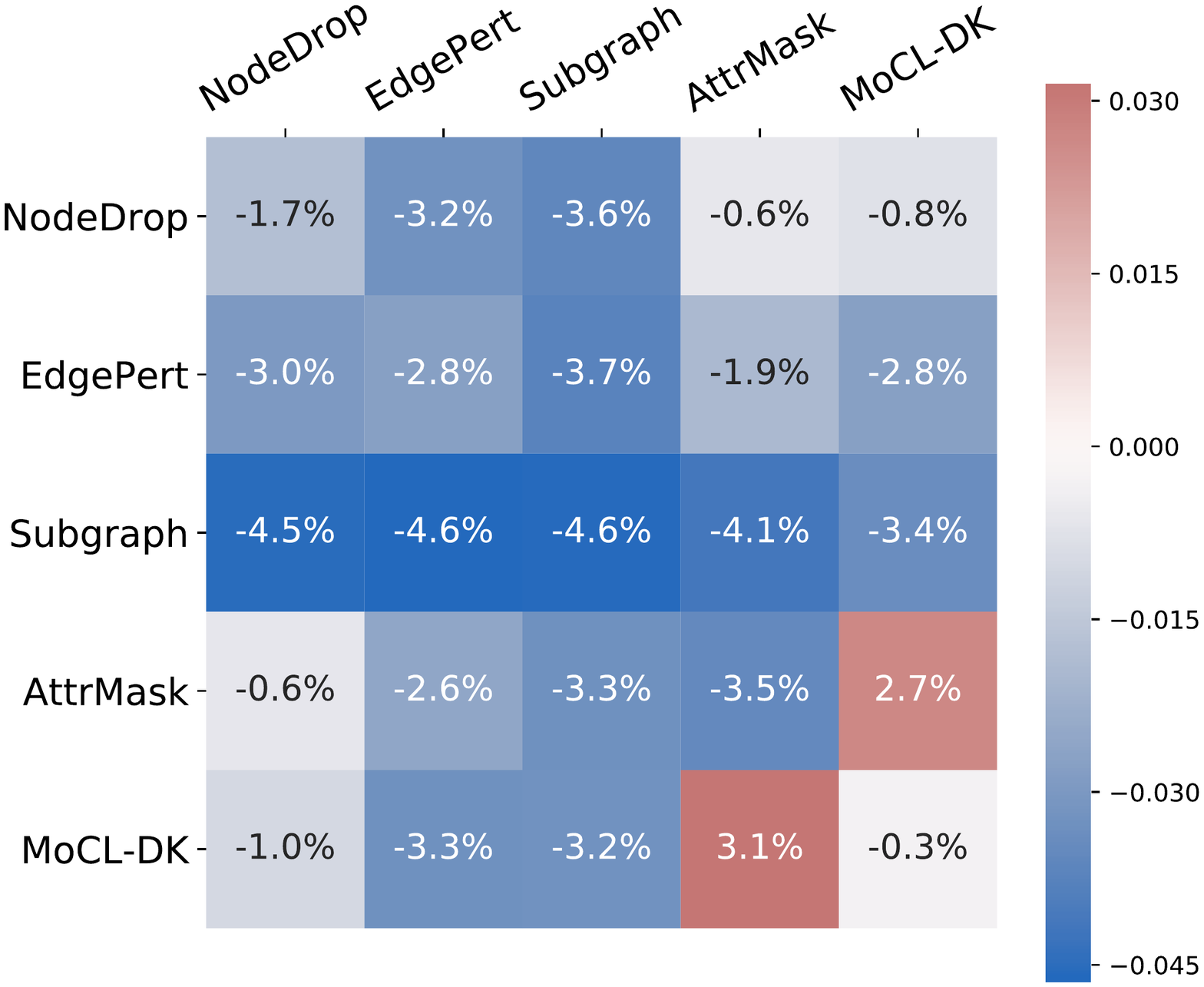}
		\caption{bbbp}
		\label{d-2}
	\end{subfigure}
	\begin{subfigure}[b]{0.32\textwidth}
		\includegraphics[width=\textwidth, trim=0 5.5cm 0 5cm, clip]{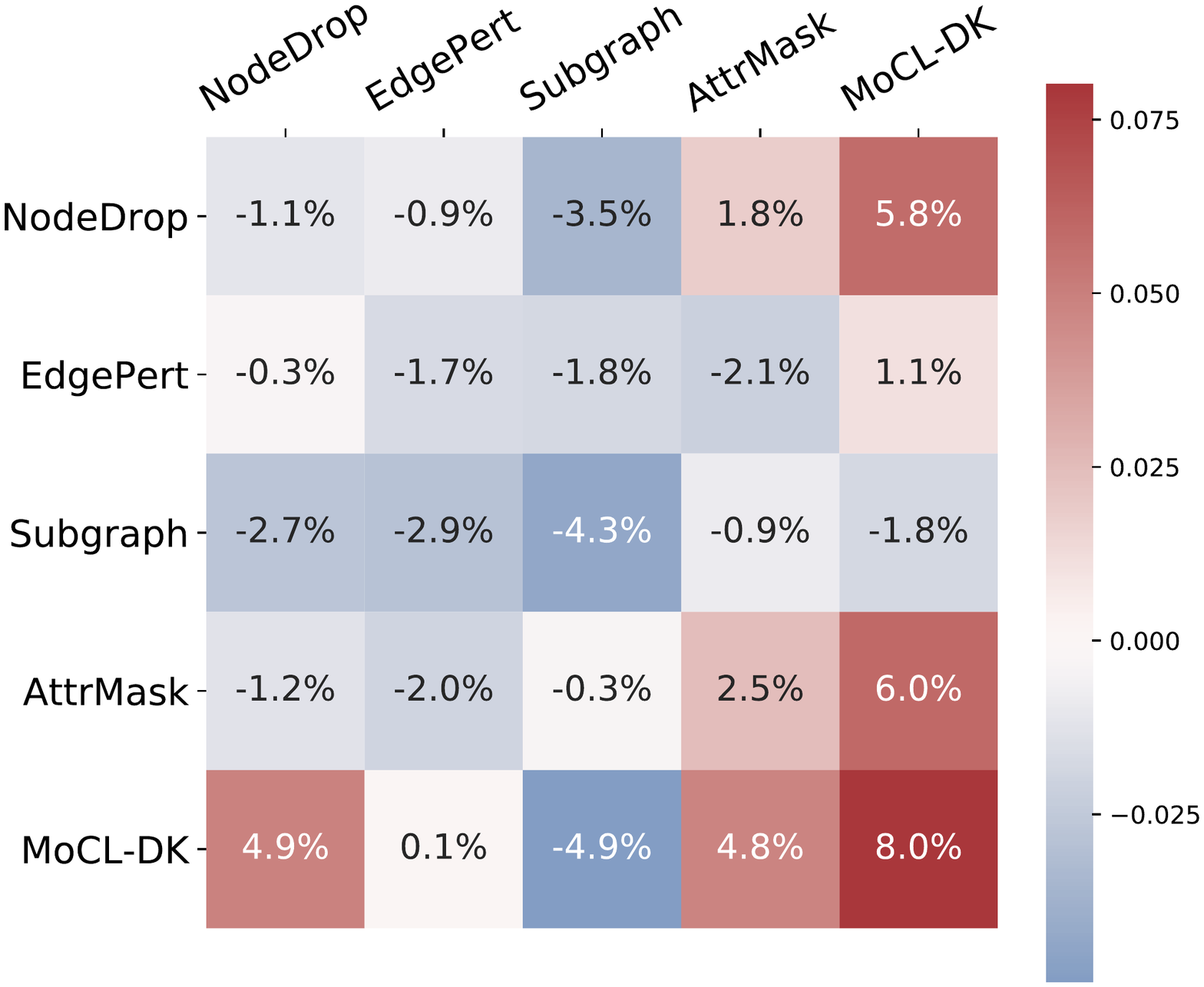}
		\caption{clintox}
		\label{d-3}
	\end{subfigure}
	\hfil
	\begin{subfigure}[b]{0.32\textwidth}
		\includegraphics[width=\textwidth, trim=0 5.5cm 0 5cm, clip]{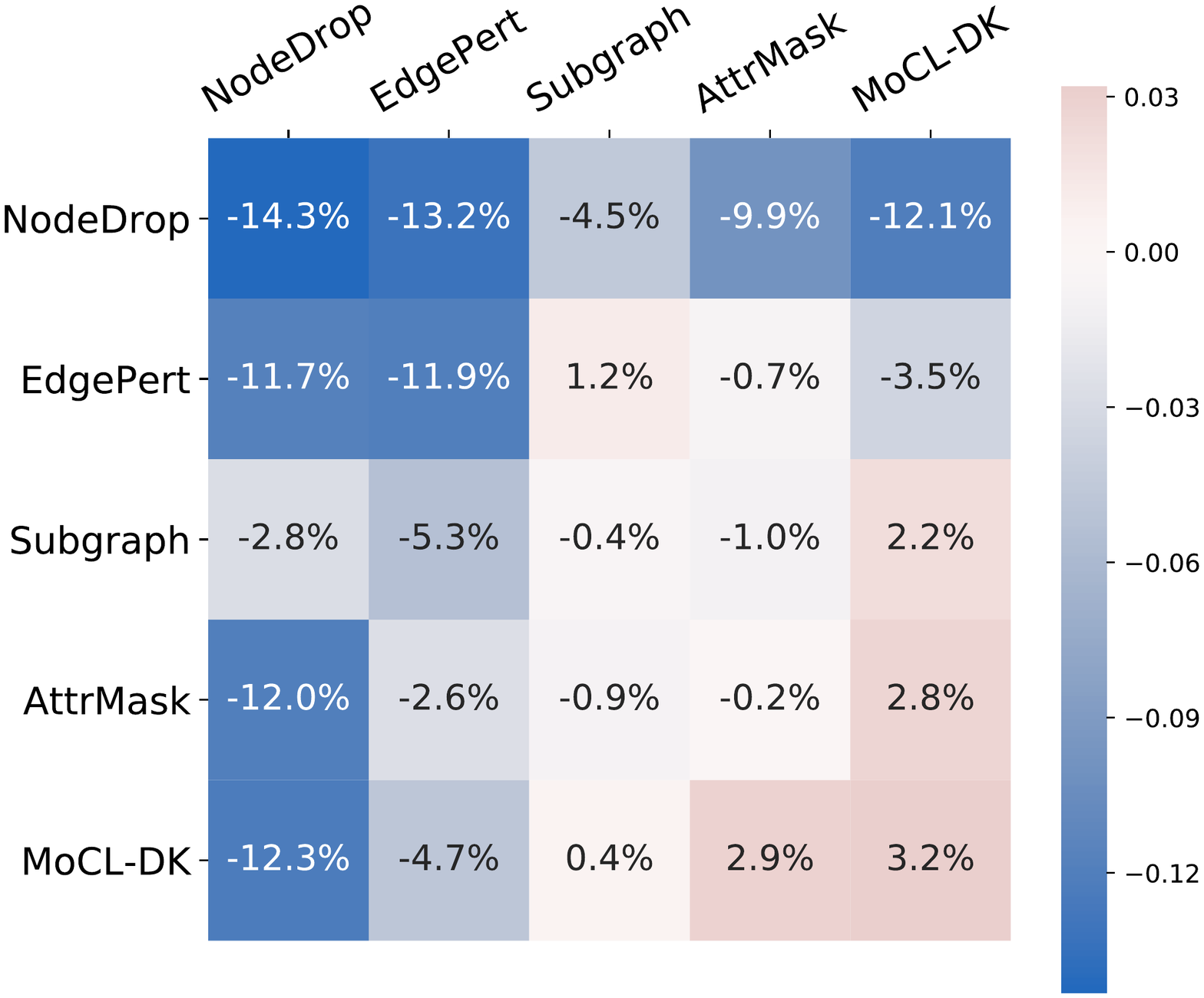}
		\caption{mutag}
		\label{d-4}
	\end{subfigure}
	\begin{subfigure}[b]{0.32\textwidth}
		\includegraphics[width=\textwidth, trim=0 5.5cm 0 5cm, clip]{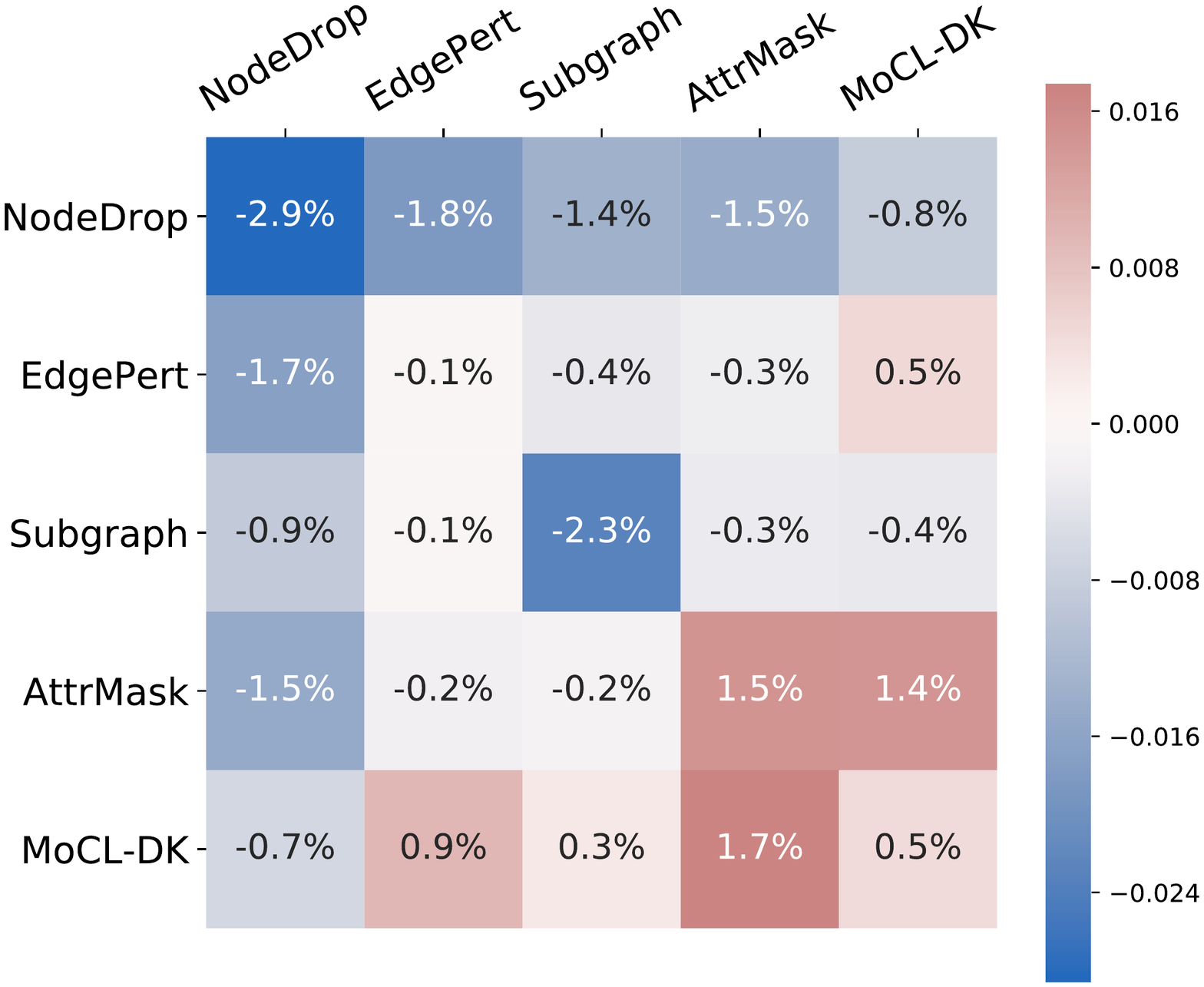}
		\caption{sider}
		\label{d-5}
	\end{subfigure}
	\begin{subfigure}[b]{0.32\textwidth}
		\includegraphics[width=\textwidth, trim=0 5.5cm 0 5cm, clip]{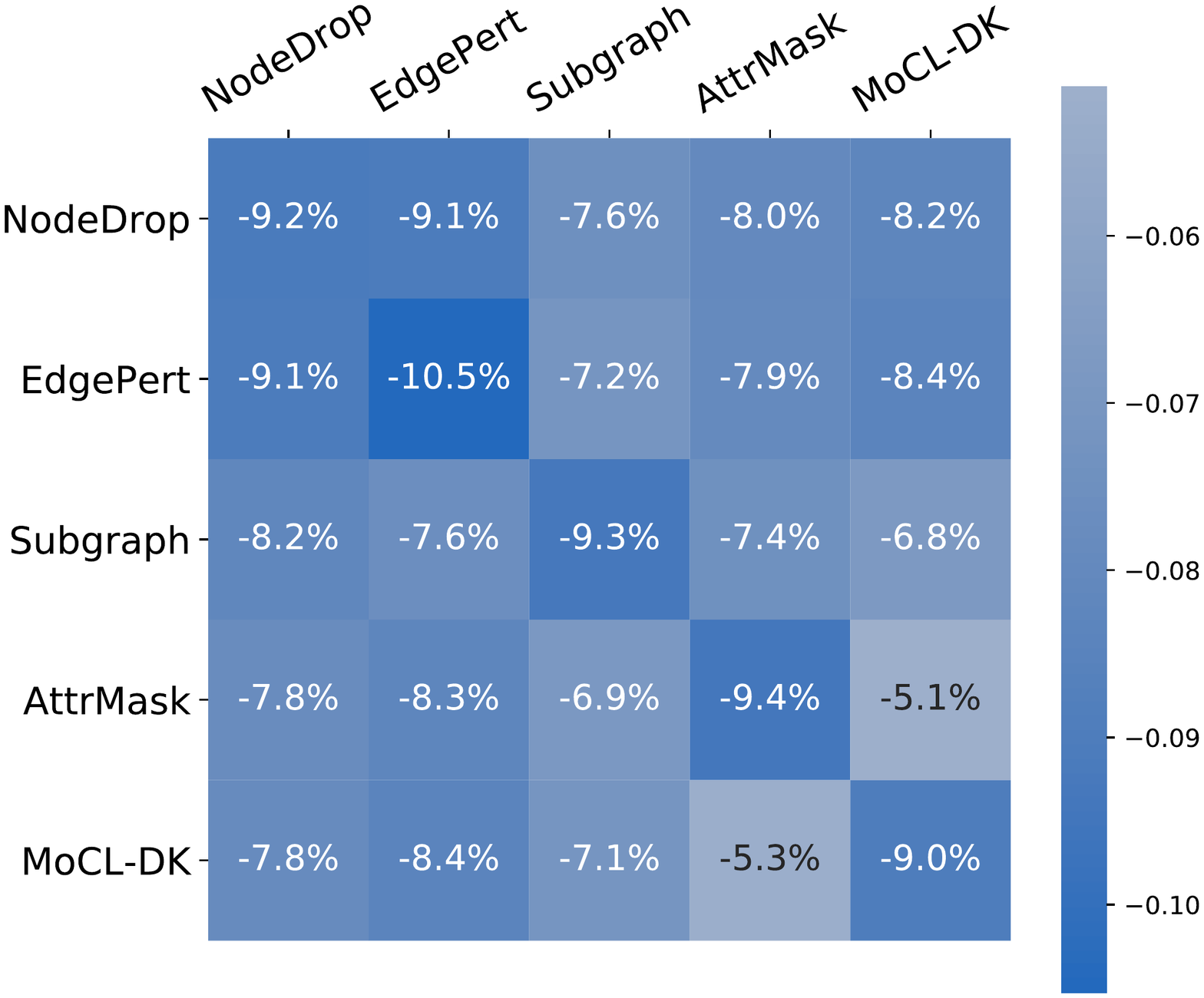}
		\caption{tox21}
		\label{d-6}
	\end{subfigure}
	\vspace{-0.8em}
	\caption{Augmentation combination under linear evaluation protocol. Each cell represents the performance difference between i) a vanilla GNN trained from scratch (upper-bound) and ii) learned representations (fixed) from the pretrained model plus a linear classifier, under a given augmentation combination. Each number is averaged from 5 runs. Blue represents negative value and red positive. Higher value is better. MoCL-DK is the proposed augmentation with local-level domain knowledge.}\label{q1-comb}
	\vspace{-0.1in}
\end{figure*}

\subsection{Evaluation Protocols}
The evaluation process follows two steps. We first pretrain a model based on any comparison method, and then evaluate the learned model on downstream tasks. We adopt two evaluation protocols:
\begin{itemize}
	\item Linear protocol: fix the representation from pretrained model and finetune a linear classifier on top of it.
	\item Semi-supervised protocol: sample a small set of labels of the downstream task and use the weights of learned graph encoder as initialization meanwhile finetune all the layers. 
\end{itemize}
which are most commonly used in literature \cite{sun2019infograph, hu2019strategies, you2020graph, zhu2020graph}. 

\subsection{Experimental Setup}
\noindent\textbf{Datasets and Features.} We use 7 benchmark molecular datasets in the literature \cite{hu2019strategies, you2020graph, sun2019infograph} to perform the experiments, which covers a wide range of molecular tasks such as binding affinity, response in bioassays, toxicity and adverse reactions:
\begin{itemize}
	\item bace \cite{subramanian2016computational}: a dataset containing the binding results between molecules and human proteins .
	\item bbbp \cite{martins2012bayesian}: a dataset measuring the blood-brain barrier penetration property of molecules.
	\item mutag \cite{rossi2015network}: a dataset recording the mutagenic effect of a molecule on a specific gram negative bacterium.
	\item clintox \& tox21 \& toxcast \cite{gayvert2016data, novick2013sweetlead, richard2016toxcast}: datasets that contains the molecule toxicity from FDA clinical trials (clintox) and in vitro high-throughput screening (tox21 and toxcast).
	\item sider \cite{kuhn2016sider}: a dataset containing the adverse drug reactions (ADR) of FDA approved drugs.
\end{itemize} 
The basic statistics of the datasets (size, tasks, molecule statistics) are summarized in Table \ref{exp-stats}. In this paper, we mainly focus on classification tasks as prior works \cite{hu2019strategies, you2020graph, sun2019infograph}, therefore we use AUC \cite{tai1994mathematical} as the major evaluation metric.  

For molecular graphs, we use both atom features and bond features as inputs. We use i) atomic number and ii) chirality tag as features for atoms and i) bond type and ii) bond directions as features for chemical bonds \cite{hu2019strategies}. 

\noindent\textbf{Model Architectures.} We use GIN \cite{xu2018powerful} as our graph encoder $f$ which has been shown to be the most expressive graph neural network layer in prior works \cite{hu2019strategies}. It also allows us to incorporate edge features of molecules into the learning process. The update rule for each GIN layer can be written as:
\vspace{-0.5em}
\begin{align*}
\mathbf{x}_i^{l+1} = \text{MLP}_{\theta}\left(\mathbf{x}_i^l + \sum\nolimits_{j\in \mathcal{N}_i}\text{ReLU}\left(\mathbf{x}_j^l + \mathbf{e}_{j, i}\right)\right),
\end{align*}
where $\mathbf{x}_i^l$ is the node representation at $l$-th layer, $\mathcal{N}_i$ denotes the neighbor nodes of $i$-th node and $\mathbf{e}_{j, i}$ represents the edge feature between node $i$ and $j$. MLP$_\theta$ is a two-layer perceptron parameterized by $\theta$. Note that MLP here is for a single GIN layer in order to make the GIN layer the most expressive. After obtaining the node representations for all atoms in a molecule, we average them to get the graph representation $\mathbf{h}$.

We use another two-layer perceptron for the projection head $g$ in our framework following literature \cite{chen2020simple, you2020graph}. It has been shown that a projection head with nonlinear transformation is necessary for a better representation of the layer before it due to information loss in the contrastive learning loss \cite{chen2020simple}. After adding a projection head, the representations at previous layer, ie., $\mathbf{h}$, can benefit more for downstream tasks. We use cosine similarity for the critic function $s(\mathbf{z}_i, \mathbf{z}_j)=\mathbf{z}_i^T \mathbf{z}_j/\|\mathbf{z}_i\|\|\mathbf{z}_j\|$ \cite{you2020graph}.

\noindent\textbf{Baselines.} For both linear and semi-supervised evaluation protocols, we adopt three types of baselines for comparison:
\begin{itemize}
	\item Vanilla GNN (\underline{Scratch}): train a standard nonlinear GNN model on labeled data of the downstream task.
	\item General GNN self-supervised learning or pretraining baselines: i) \underline{InfoGraph} \cite{sun2019infograph}, which maximizes the mutual information between nodes and graph; ii) \underline{Edge Pred} \& \underline{Context} \underline{Pred} \cite{hu2019strategies}: which uses the node embeddings to predict graph edge and neighbor context in order to learn meaningful node representations; iii) \underline{Masking} \cite{hu2019strategies}: which masks the atom attributes and tries to predict them.
	\item Graph contrastive learning baselines: we adopt the four types of general augmentations for graph in \cite{you2020graph}: i) \underline{node dropping}; ii) \underline{edge perturbation}; iii) \underline{subgraph extraction}; iv) \underline{attribute} \underline{masking} for comparison. We also add linear procotol resutls reported in \underline{MICRO-Graph} \cite{zhang2020motif} which is a motif-based contrastive method for comparison (no public code available).
\end{itemize}

\begin{table*}[tb]
	\small \centering
	\setlength\tabcolsep{3.6pt} % default value: 6pt
	\begin{tabular}{c|ccccccc|ccccccc}
		\hhline{===============}
		Protocol & \multicolumn{7}{c|}{Linear   Protocol}                    & \multicolumn{7}{c}{Semi-supervised   Protocol}            \\ \hline
		Method $|$ Dataset & bace  & bbbp  & clintox & mutag & sider & tox21 & toxcast & bace  & bbbp  & clintox & mutag & sider & tox21 & toxcast \\ \hhline{=|=======|=======}
		scratch          		& 0.785 & 0.861 & 0.647   & 0.918 & 0.606 & 0.820 & 0.710   & 0.525 & 0.695 & 0.494   & 0.803 & 0.552 & 0.670 & 0.530   \\ \hline
		InfoGraph     		& 0.594 & 0.611 & 0.458   & 0.771 & 0.502 & 0.615 & 0.562   & 0.614 & 0.735 & 0.487   & 0.887 & 0.523 & 0.589 & 0.535   \\
		contextpred      		& 0.522 & 0.724 & 0.506   & 0.819 & 0.498 & 0.554 & 0.542   & 0.566 & 0.731 & 0.502   & 0.846 & 0.525 & 0.659 & 0.514   \\
		edgepred         		& 0.662 & 0.592 & 0.504   & 0.622 & 0.502 & 0.500 & 0.501   & 0.604 & 0.694 & 0.486   & 0.915 & 0.545 & 0.615 & 0.529   \\
		masking          		& 0.678 & 0.764 & 0.581   & 0.826 & 0.566 & 0.722 & 0.617   & 0.621 & \textbf{0.776} & 0.585   & 0.879 & \textbf{0.551} & 0.640 & 0.538   \\ \hline
		drop\_node       		& 0.746 & 0.843 & 0.635   & 0.775 & 0.577 & 0.728 & 0.633   & 0.603 & 0.767 & 0.492   & 0.836 & 0.542 & 0.656 & 0.525   \\
		perturb\_edge    		& 0.657 & 0.833 & 0.630   & 0.799 & 0.605 & 0.715 & 0.619   & 0.527 & 0.748 & 0.516   & \textbf{0.938} & 0.547 & 0.629 & 0.516   \\
		subgraph         		& 0.629 & 0.815 & 0.603   & 0.914 & 0.583 & 0.727 & 0.625   & 0.565 & 0.769 & 0.539   & 0.918 & 0.548 & 0.656 & 0.514   \\
		mask\_attributes 		& 0.796 & 0.826 & 0.671   & 0.916 & 0.621 & 0.726 & 0.623   & 0.622 & 0.710 & 0.478   & 0.897 & 0.549 & \textbf{0.666} & 0.543   \\
		MICRO-Graph				& 0.708 & 0.830 & \textbf{0.735}   & -	  & 0.573 & 0.718 & 0.595   & -		& - 	& - 	  & -	 & -	& - 	& - \\ \hline
		MoCL-DK          		& 0.801 & 0.870 & 0.727   & \textbf{0.950} & 0.615 & 0.740 & 0.636   & \textbf{0.650} & 0.765 & \textbf{0.588}   & 0.903 & 0.546 & 0.645 & \textbf{0.539}   \\
		MoCL+AttrMask 		& \textbf{0.831} & \textbf{0.892} & 0.695   & 0.947 & \textbf{0.623} & \textbf{0.768} & \textbf{0.653}	 & 0.630 & 0.748 & 0.549   & 0.909 & 0.536 & 0.661 & 0.536   \\ \hline
		MoCL-DK-G(LS)    		& 0.831 & 0.892 & 0.724   & 0.958 & 0.623 & \textbf{0.777*} & \textbf{0.659*}	 & 0.662 & 0.766 & 0.623   & 0.907 & 0.558 & 0.666 & \textbf{0.547*}   \\ 
		MoCL-DK-G(CL)    		& \textbf{0.845*} & 0.905 & \textbf{0.750*}   & \textbf{0.969*} & \textbf{0.628*} & 0.768 & 0.653	 & \textbf{0.706*} & \textbf{0.809*} & \textbf{0.623*}   & 0.916 & 0.565 & 0.686 & 0.546  \\ 
		MoCL+AttrMask-G(CL)  & 0.833	& \textbf{0.911*}	& 0.747	  & 0.962 & 0.625 & 0.774 & 0.654	& 0.695	& 0.806	& 0.618	 & 0.913   & \textbf{0.567*} & \textbf{0.687*} & 0.544   \\ \hhline{=|=======|=======}
	\end{tabular}
	\caption{Averaged test AUC of comparison methods under linear and semi-supervised protocol (5 runs). \textbf{Bold} number denotes the best performance for local-level (augmentation) comparison. \textbf{Bold*} number denotes the best performance after incorporating global similarity information (MoCL-G). LS and CL represents least-square and contrastive global loss, respectively.}\label{final}
	\vspace{-0.3in}
\end{table*}

\noindent\textbf{Implementation Details.} We use 3 layers of GIN for all methods since 3-hops neighborhood covers most aromatic rings and is usually sufficient for molecular structure learning \cite{rogers2010extended}. The dimensions for GIN layer and embedding layer are 512 and 128 respectively. We use Adam as optimizer with initial learning rate of 0.001 for all methods. We use dropout ratio 0.5 for GIN layers and default settings for baselines. The batch size is 32 across all scenarios. For pretraining models, the running epoch is fixed to 100. For downstream tasks, we use early stop via validation set. We implement all models using Pytorch \cite{paszke2019pytorch} and run them on Tesla K80 GPUs. 

The variation of results for a dataset comes from two sources, the pretrained model and the downstream task. By comparing them, we find the variation of pretrained model (by applying different seeds) is much smaller than the variation of downstream task (by different training-testing splits). Therefore, for each dataset, we use its molecular graphs to pretrain a model (1 seed) and then apply it to downstream task on the same dataset using different splits (5 seeds). We do not evaluate transfer learning setting in this paper where a pretrained model is applied to another dataset. During downstream task, we split the dataset into training (0.8), validation (0.1) and testing (0.1) set, we use validation set for early stop and evaluate the AUC on testing set. For semi-supervised protocol where only a small fraction of labels is used to train, since the data sizes are different, the ratio is picked from \{0.01, 0.05, 0.5\} such that around 100 molecules being selected for each dataset. For local-level domain knowledge, we use augmentation ratio $0.2$ for general augmentations as prior work \cite{you2020graph} and different augmentation times $\{1, 2, 3, 5\}$ for the proposed method. For example, MoCL-DK3 denotes applying domain augmentation 3 times. For global-level domain knowledge part, we try $\lambda=\{0.5, 1.0, 5.0, 10.0\}$ and 4 different nearest neighbor sizes for each dataset based on its size. We use ECFP with dimension 1024 to calculate the global similarity. The complete implementation details can be found in Appendix.

\vspace{-1em}
\subsection{Local-level domain knowledge (\textbf{Q1})}
We first examine whether the proposed augmentation helps learn a better representation. Since the contrastive framework involves two correlated views, different augmentation schemes can be applied to each view. Figure \ref{q1-comb} shows the results of different augmentation combinations under \underline{linear} protocol for all datasets (the results of toxcast is similar as tox21 therefore we remove it due to space limit). MoCL-DK represent applying domain augmentation by only once. We can see that i) the representations from MoCL-DK (diagonals) plus a linear classifier yield prediction accuracies which are on-par with a deep learning model train from scratch (bace, bbbp, sider), or even better than it (clintox, mutag). ii) the proposed augmentation MoCL-DK combined with other augmentations almost always produce better results compared to other combinations (rows and columns that contain MoCL-DK are usually higher). iii) Attribute masking and MoCL-DK are generally effective across all scenarios, combining them often yields even better performance. This verifies our previous assumption that MoCL-DK and attribute masking does not violate the biological assumption and thus works better than other augmentations. Moreover, harder contrast, e.g., combination of different augmentation schemes benefits more as compared to one augmentation schemes (MoCL-DK + AttrMask often produce the best results). This phenomenon is reasonable and also observed in prior works \cite{you2020graph}. 

\begin{figure}[tb]
	\centering
	\begin{subfigure}[b]{0.23\textwidth}
		\includegraphics[width=\textwidth, trim=0 5cm 0 4cm, clip]{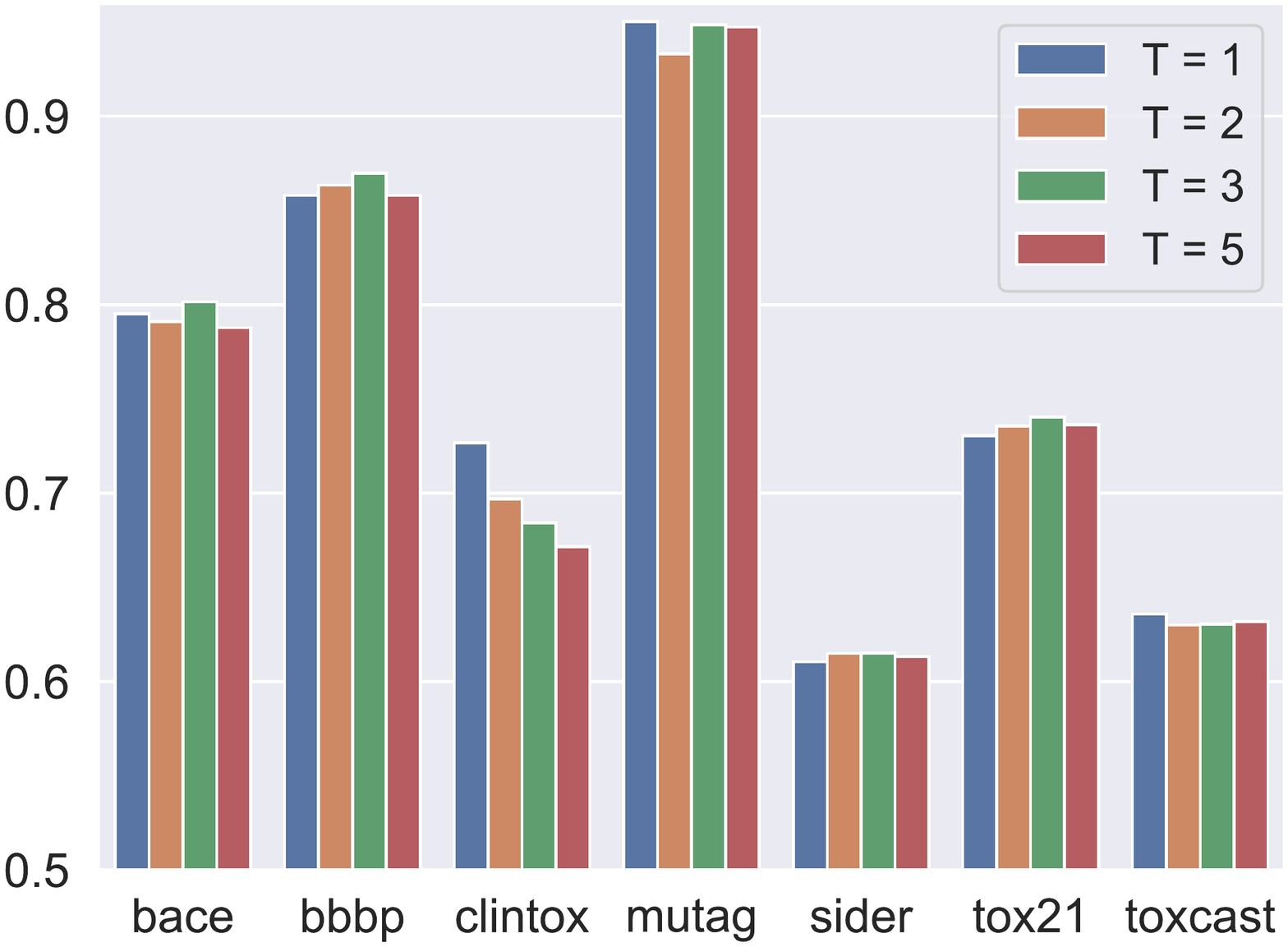}
		\caption{Linear protocol}
		\label{q1-mocl-1}
	\end{subfigure}
	\begin{subfigure}[b]{0.23\textwidth}
		\includegraphics[width=\textwidth, trim=0 5cm 0 4cm, clip]{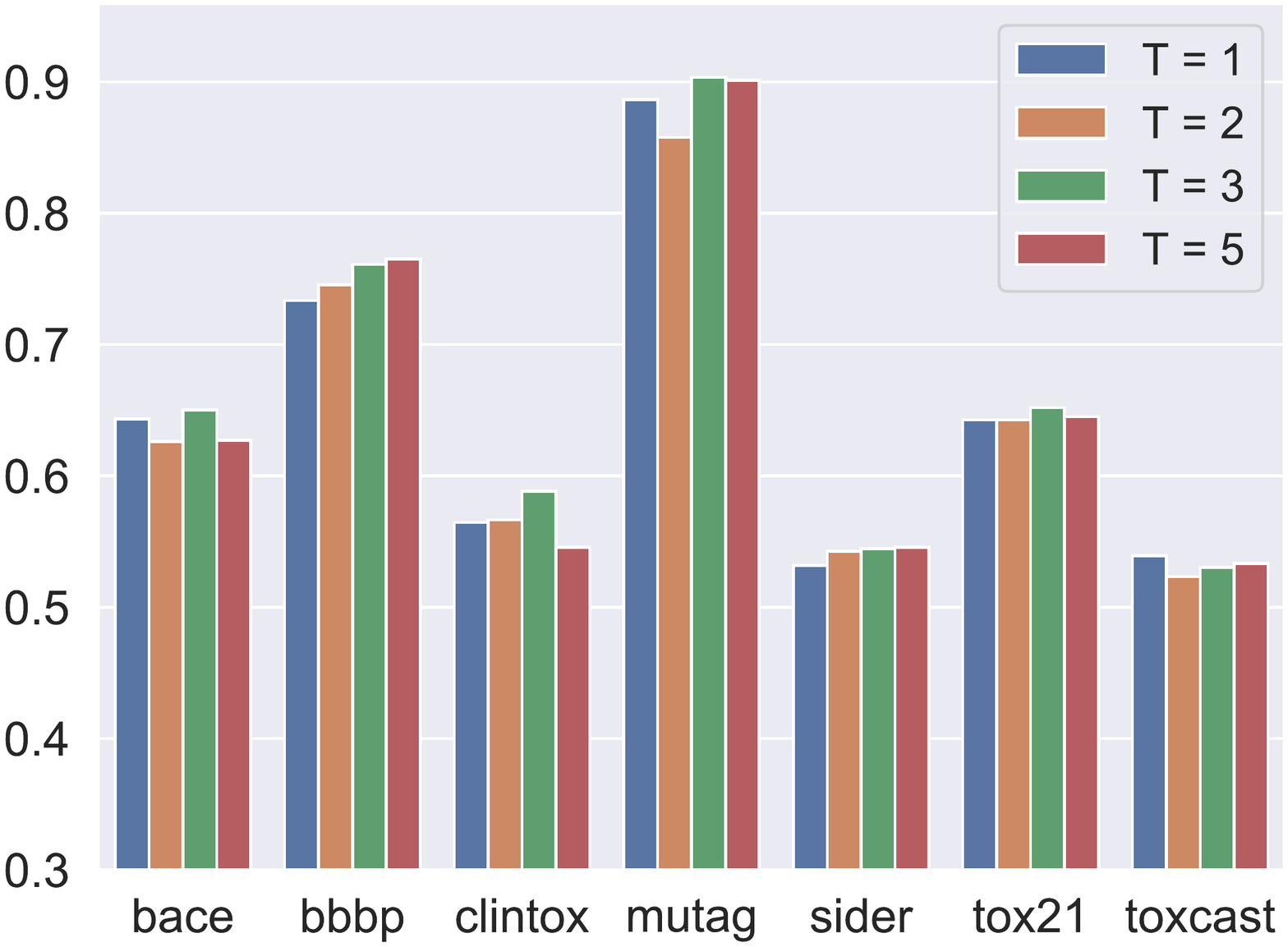}
		\caption{Semi-supervised protocol}
		\label{q1-mocl-2}
	\end{subfigure}
	\vspace{-0.3em}
	\caption{Average test AUC of MoCL-Local across different augmentation strengths (repeat times) for all datasets.}\label{q1-mocl}
	\vspace{-0.2in}
\end{figure}

For semi-supervised protocol, the results are weaker, we did not include the augmentation combination figure due to space limit. But the complete results for all comparison methods for both linear and semi-supervised protocol can be found in Table \ref{final}, where the next-to-bottom panel represents results for proposed augmentation and the bottom panel presents global results which we will mention in the next subsection.

The proposed augmentation MoCL-DK can be applied multiple times to generate more complicated views. We tried over a range of different augmentation strengths and report the corresponding results for all datasets in Figure \ref{q1-mocl}. We can see that for most datasets, as we apply more times the proposed augmentation, the performance first increases and then decreases. MoCL-DK3 usually achieves better results than others. For certain datasets (clintox, toxcast) the trend is not very clear between the two evaluation protocols.

\subsection{Global-level domain knowledge (\textbf{Q2})}
We next study the role of global-level domain knowledge by examining the following sub-questions: i) Does global similarity helps general (baseline) augmentations? Does it helps the proposed augmentation? Are the effectiveness the same? ii) How do different global losses behave, i.e., direct supervision as least square loss v.s. contrastive loss, across all datasets, which one is better?

\begin{figure}[tb]
	\centering
	\begin{subfigure}[b]{0.23\textwidth}
		\includegraphics[width=\textwidth, trim=0 6.5cm 0 6cm, clip]{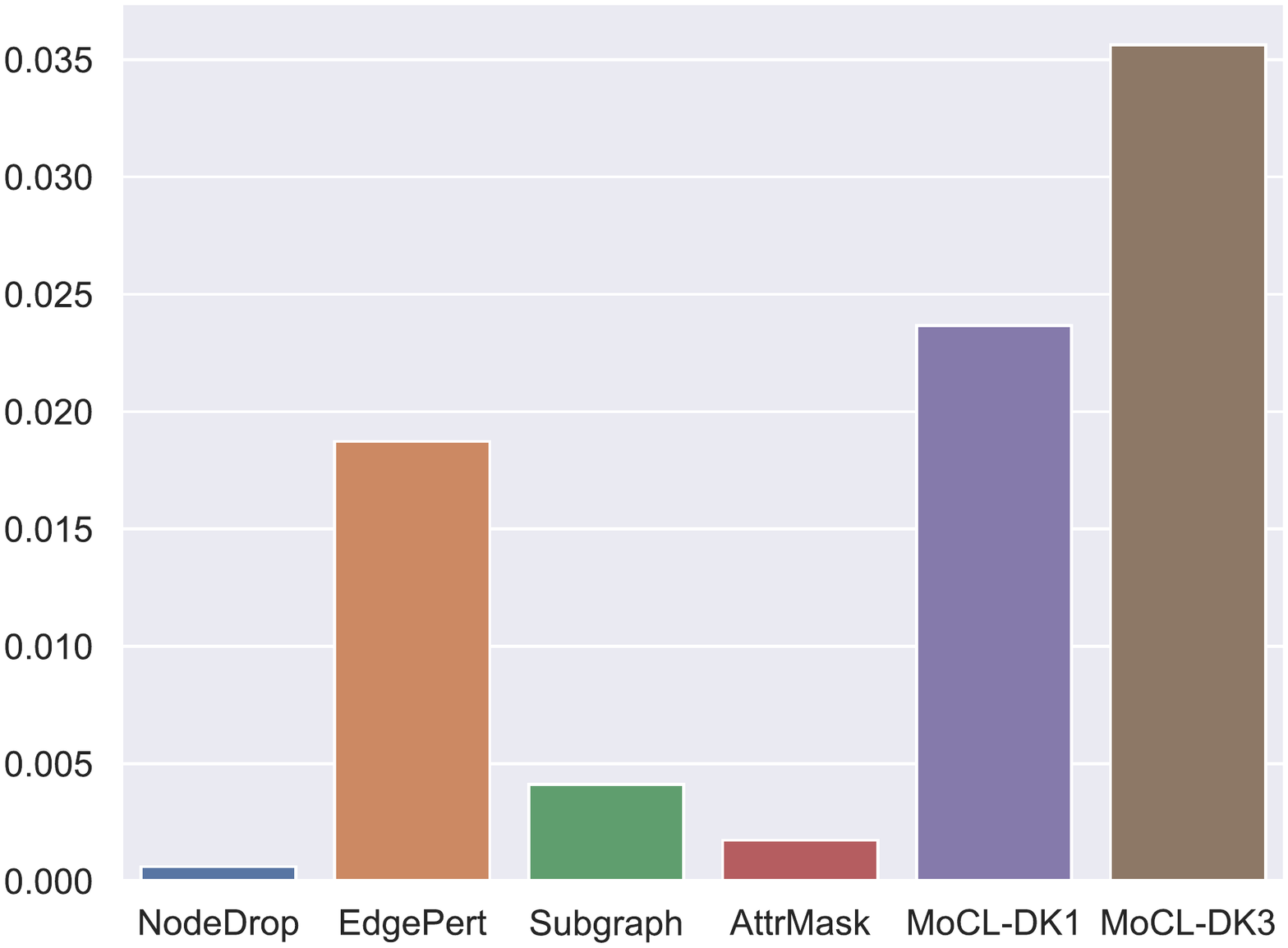}
		\caption{Linear protocol}
		\label{q2-linear}
	\end{subfigure}
	\begin{subfigure}[b]{0.23\textwidth}
		\includegraphics[width=\textwidth, trim=0 6.5cm 0 6cm, clip]{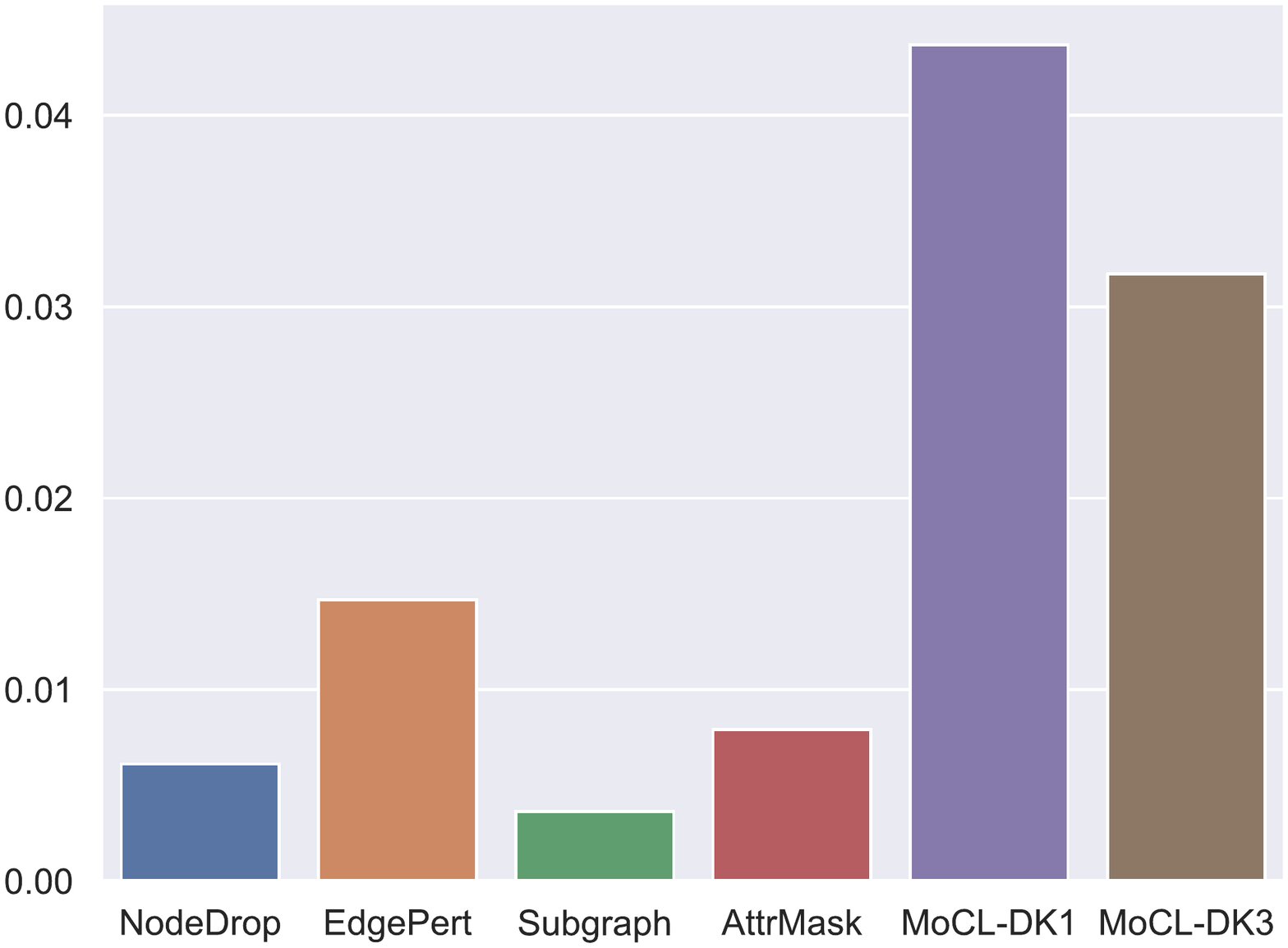}
		\caption{Semi-supervised protocol}
		\label{q2-semi}
	\end{subfigure}
	\vspace{-0.3em}
	\caption{Average test AUC \underline{gain} from global domain knowledge for different augmentations across all datasets.}\label{q2-gain}
\end{figure}

Figure \ref{q2-gain} shows the performance gain by incorporating global similarity information for general (baseline) augmentations and the proposed augmentation. Each bar represents the median gain across all 7 datasets for a particular augmentation scheme. We can see that global information generally improves all augmentation schemes (the bars are positive). Interestingly, the gain for proposed domain augmentation (MoCL-DK1 and MoCL-DK3) are much higher as compared to other augmentations schemes. Note that we used the same set of global-level hyper-parameters for all augmentations for fair comparison. Table \ref{q2-2} shows the performance for different global losses under both evaluation protocols. We can see that contrastive loss (CL) for the global similarity achieves better results than directly using it as supervision by least-square loss (LS). 

We summarize the complete results for all comparison methods in Table \ref{final}. We can see that i) contrastive learning works generally better than traditional graph pretraining methods, especially in linear protocol; ii) The proposed augmentation outperforms general augmentations. By combining MoCL augmentation and attribute masking, the results are even better for some datasets; iii) The global similarity information further improves the learned representations. Moreover, without combining with attribute masking, MoCL augmentation only already achieves the best performance under most scenarios after adding global information. The learned representations plus a linear classifier can achieve higher accuracy than a well-trained deep learning model. In summary, the proposed method is demonstrated to be effective for various molecular tasks.

\begin{table}[tb]
	\begin{tabular}{ccccc}
		\hline
		Protocol & \multicolumn{2}{c}{Linear} & \multicolumn{2}{c}{Semi-supervised} \\ \hline
		Dataset  & LS       & CL          & LS            & CL              \\ \hline
		bace     & 0.831        & \textbf{0.845}       & 0.662             & \textbf{0.701}           \\
		bbbp     & 0.891        & \textbf{0.903}       & 0.766             & \textbf{0.809}           \\
		clintox  & 0.724        & \textbf{0.750}       & 0.608             & \textbf{0.619}           \\
		mutag    & 0.954        & \textbf{0.963}       & 0.895             & \textbf{0.907}           \\
		clintox  & 0.623        & \textbf{0.628}       & 0.551             & \textbf{0.563}           \\
		tox21    & 0.774        & 0.768       & 0.655             & \textbf{0.686}           \\
		toxcast  & 0.659        & 0.653       & 0.547             & 0.546           \\ \hline
	\end{tabular}
	\caption{Comparison between different global losses under MoCL-DK1 augmentation. LS: directly using global similarity and optimize by least-square loss; CL: contrastive loss using nearest neighbor derived from global similarity.}\label{q2-2} 
	\vspace{-0.4in}
\end{table}

\subsection{Sensitivity Analysis (\textbf{Q3})}
\begin{figure}[tb]
	\centering
		\includegraphics[width=0.46\textwidth, trim=2cm 6.5cm 0cm 7cm, clip]{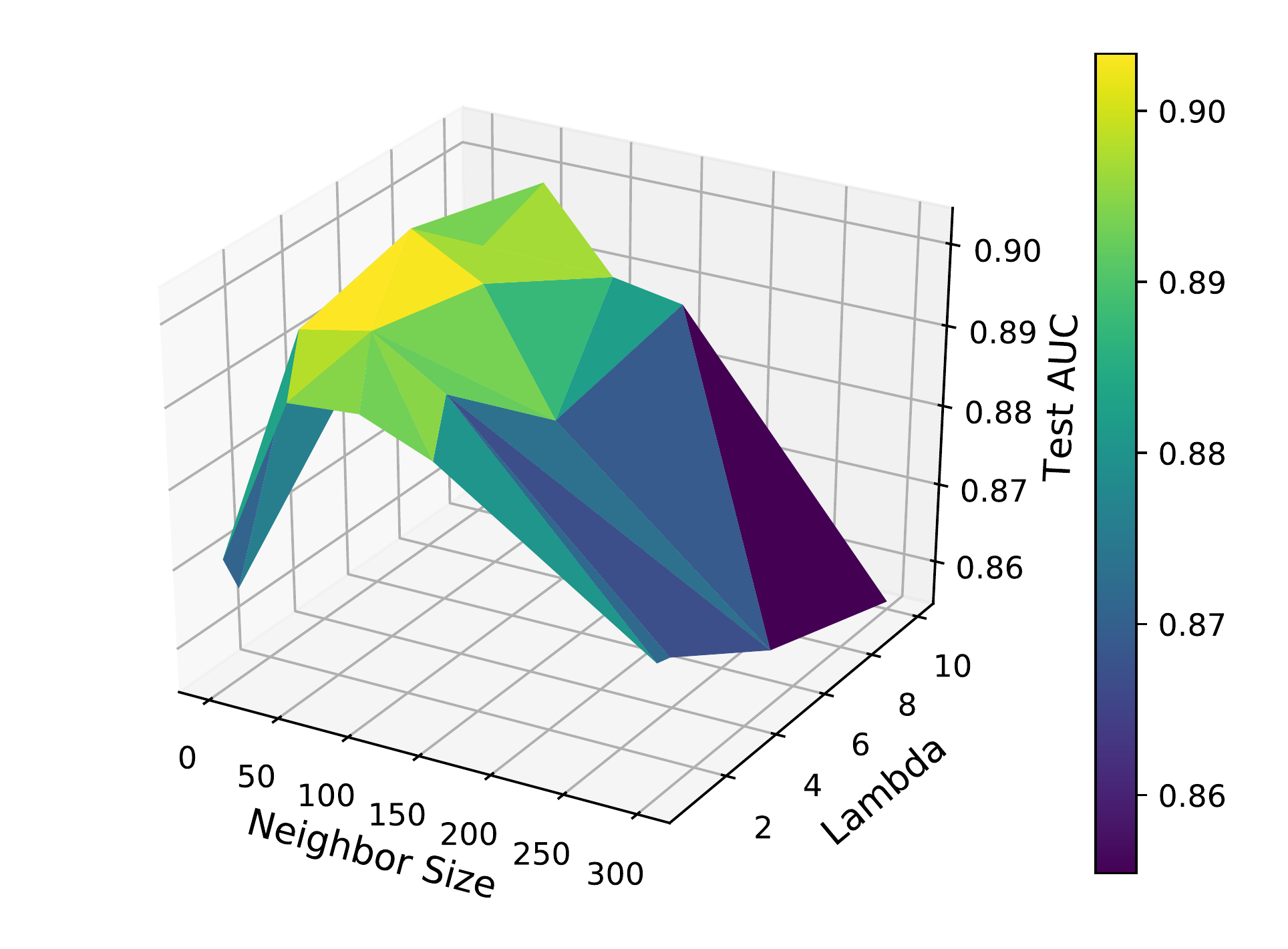}
	\caption{Average test AUC of different neighbor size and $\lambda$ for MoCL-DK1-G under linear protocol (dataset: bbbp).}\label{q3}
\end{figure}

Finally we check the sensitivity of global-level hyper-parameters, ie., the neighbor size and $\lambda$ that controls the weight between local and global loss. Figure \ref{q3} shows the performance surface under different hyper-parameter combinations of the proposed method for bbbp dataset. We can see that a relatively smaller neighbor size (not too small) and larger weights (not too large) for the global loss leads to a best result. Other datasets also show the similar pattern.  

\vspace{-0.5em}

\subsection{Discussion}
We provide additional observations and discussion in this subsection. First, we observe that representations which perform well under linear evaluation do not guarantee to be better in the semi-supervised setting. Since we finetune all the layers in semi-supervised learning, an overly delicate representation as initialization may not produce the best results in a fully nonlinear setting. Second, the effectiveness of contrastive learning also depends on the property of the dataset as well as the nature of the task. For example, single property prediction (mutag, bbbp) benefits more from pretraining as compared to toxicity prediction (tox21, toxcast) since it depends not only on the compound structure, but also the cellular environment. Therefore, incorporating drug-target network information and system biology data may be more helpful to these datasets, which is our future direction. %In the future, we may also take into account activity cliff where a small change of chemical structure leads to a large difference of biological activity.